\documentclass[useAMS]{mn2e}
\usepackage{amsmath}
\usepackage{amsfonts}
\usepackage{epsfig}
\usepackage{graphicx}
\numberwithin{equation}{section}

\newcommand{\p}{\partial}
\newcommand{\case}{\textstyle\frac}
\newcommand{\bv}{{\mbox{\boldmath $v$}}}
\newcommand{\br}{{\mbox{\boldmath $r$}}}

\newcommand{\eps}{\epsilon}
\newcommand{\ccase}{\textstyle\frac}

\newcommand{\ba}{\begin{array}}
\newcommand{\ea}{\end{array}}
\def\({\left(}\def\){\right)}
\def\il{\int\limits}
\def\Im{\textrm{Im}\,}
\title[Effect
of Angular Momentum Distribution on Loss-Cone Instability]{Effect of Angular
Momentum Distribution
\\on Gravitational Loss-Cone Instability\\ in Stellar Clusters Around
Massive BH}
\author[E. V. Polyachenko et al.]
       {E.~V.~Polyachenko,$^1$\thanks{E-mail: epolyach@inasan.ru}, V. L. Polyachenko,$^1$
        I. G. Shukhman,$^2$\thanks{E-mail: shukhman@iszf.irk.ru}\\
       $^1$Institute of Astronomy, Russian Academy of Sciences, 48 Pyatnitskya St., Moscow 119017, Russia\\
       $^2$Institute of Solar-Terrestrial Physics, Russian Academy of Sciences, Siberian Branch, P.O. Box 291, Irkutsk 664033, Russia}

\date{Accepted \qquad
      Received }

\pagerange{\pageref{firstpage}--\pageref{lastpage}} \pubyear{2006}

\begin{document}
\maketitle

\label{firstpage}

\begin{abstract}
\small Small perturbations in spherical and thin disk stellar
clusters surrounding massive a black hole are studied. Due to the
black hole, stars with sufficiently low angular momentum escape
from the system through the loss cone. We show that stability
properties of spherical clusters crucially depend on whether the
distribution of stars is monotonic or non-monotonic in angular
momentum. It turns out that only non-monotonic distributions can
be unstable. At the same time the instability in disk clusters is
possible for both types of distributions.
\end{abstract}

\section{Introduction}

The study of the gravitational loss-cone instability, a far analog of the plasma
cone instability, has begun with the work of V. Polyachenko (1991),  in which a
simplest analytical model of thin disk stellar cluster has been treated. The
interest to the problem of stability of stellar clusters has been revived
recently by detailed investigation by Tremaine (2005) and Polyachenko,
Polyachenko, Shukhman, (2007; henceforth, Paper I) of low mass clusters around
massive black holes. The both papers have considered stability of small amplitude
perturbations of stellar clusters of disk-like and spherical geometry.

Tremaine (2005) has shown using Goodman's (1988) criterion that thin disks with
symmetric DFs over angular momentum and empty loss cone are generally unstable. By
contrast, analyzing perturbations with spherical numbers $l=1$ and $l=2$, he
deduced that spherical clusters with monotonically increasing DF of angular
momentum should be generally stable.

Later we demonstrated (see Paper I) that spherical systems with
non-monotonic distributions may be unstable for sufficiently small-scale
perturbations $l \ge 3 $, while the harmonics $l=1,2$ are always stable. For the sake
of convenience, we have used two assumptions. The first one is that the Keplerian
potential of the massive black hole dominates over a self-gravitating potential
of the stellar cluster (which does not mean that one can neglect the latter).
Then the characteristic time of system evolution is of the order of the orbit
precessing time, which is slow, compared to typical dynamical (free fall) time.
Since a star makes many revolutions in its almost unaltered orbit, we can regard
it as to be ``smeared out'' along the orbit in accordance with passing time, and
study evolution of systems made of these extended objects.

The second assumption is a so called {\it spoke approximation}, in
which a system consists of near-radial orbits only. This
approximation was earlier suggested by one of the authors
(Polyachenko 1989, 1991). The spoke approximation reduces the
problem to a study of rather simple analytical characteristic
equations controlling small perturbations of stellar clusters.

There are two questions that naturally arise in this context.
First: Does the instability remain when abandoning the assumption
of strong radial elongation of orbits? Second: Does the
instability occur in spheres with monotonically increasing
distributions in angular momentum if one consider smaller-scale
perturbations with $l \ge 3$? The aim of the paper is to provide
answers to these questions.

To achieve the task we use semi-analytical approach based on
analysis of integral equations for slow modes elaborated recently
in Polyachenko (2004, 2005) for thin disks, and in Paper I for
spherical geometry. Following Paper I, we shall restrict ourselves
to studying monoenergetic models with DFs in the form
\begin{align}\label{eq:1.1}
F(E,L)=A\,\delta(E-E_0)\,f(L).
\end{align}
The models specified by function $f(L)$ are suitable for studying the effects of angular momentum distribution on gravitational loss-cone instability. On the other hand, the Dirac $\delta$-function permits one to reduce the integral equations for slow modes to one-dimensional integral equations, and to advance substantially in analytical calculations.

Several arguments can be brought in favour of our simplified approach. First of all, the Lynden-Bell derivative (see Paper I, eq. 4.7) of the DF with respect to angular momentum $L$, keeping $J = L + I_1$ constant (here $I_1$ is the radial action) in the limit where the slow mode approximation is applicable, can be replaced by a derivative, keeping energy $E$ constant:
$$
\left(\frac{\p F}{\p L} \right)_{LB} = \Omega_\textrm{pr} \left(\frac{\p  F}{\p E} \right)_L  + \left(\frac{\p F}{\p L} \right)_E \approx \left(\frac{\p F}{\p L} \right)_{E},
$$
because $\Omega_\textrm{pr}$ is small.
Thus, the derivative over energy is not included into the slow integral equation, and one can loosely say, that dependence on energy is only parametric. Another argument is that the results of independent study by Tremaine (2005), who used a non-monoenergetic DF, are in agreement with our conclusions.

Section 2 is devoted to spheres, Section 3  -- to thin disks with symmetric DFs.
The sections are organized alike. In the beginning we derive integral
equations for initial distribution functions in the form (\ref{eq:1.1}). Then
follow analytical and numerical investigations of these equations. We demonstrate
that by contrast to the case of near-Keplerian sphere, the loss-cone
instability in disks takes place even for the monotonic DF, $df/d|L|>0$, provided the
precession is retrograde and the loss cone is empty: $f(0)=0$. Sec. 2 is
complimented by stability analysis of models with circular orbits, which of
course doesn't belong to the class of monoenergetic models of (\ref{eq:1.1})
type.

In the last,  Section 4, we discuss the results and some
perspectives of further studies.

\section{Spherical systems}

\subsection {Integral equation for slow modes in monoenergetic models}

For the near-Keplerian systems, the slow integral equation, which
has been derived in our Paper I (see there Eq. (4.8)), is neatly
suited. In contrast to Paper I, we shall not assume here strong
elongation of orbits, i.e. we shall go beyond the spoke
approximation.

Since energies of all stars are identical, the unperturbed DF depends on one
variable only. It is convenient to use a dimensionless angular momentum $\alpha =
L/L_\textrm{circ}(E_0)$, where $L_\textrm{circ}$ is the angular momentum on circular orbits: $L_\textrm{circ}(E_0) = GM_c/(2|E_0|)^{1/2}$, $M_c$ is the central point
mass, $G$ is the gravitational constant. The frequency of stellar radial
oscillations $\Omega_1(E_0) = (2|E_0|)^{3/2}/(GM_c)$, and the radius of the system
$R(E_0)=GM_c/|E_0|$ are independent of the angular momentum. For shorthand
notations, we shall omit the argument $E_0$.

The normalization constant $A$ is taken so that a mass of the
spherical system surrounding the central mass is equal to $M_G$
(we assume the ratio $\eps\equiv M_G/M_c$ to be small: $\eps \ll
1$):
$$
M_G = \int F d\Gamma = 2\,(2\pi)^3 \int \frac{dE}{\Omega_1(E)}
\int\limits_0^{L_{\rm circ}} L\,dL\,F(E,L).
$$
If one defines the normalization of the dimensionless DF over angular momentum
$f$ (see (1.1)) as $\int_0^1 d\alpha \,\alpha f(\alpha) =1$, then normalization
factor $A$ in (1.1) is
\begin{align}\label{eq:1s.0}
   A = \frac{\Omega_1 M_G}{16\pi^3 L_\textrm{circ}^2}.
\end{align}
It allows to represent the kernel of the integral equation
(formula (4.8) in the Paper I) in the form
$$
P^{(l)}_{s,\,s'}(E,L; E', L') = \frac{8\,\pi^2\,(2l+1)}{R}\,\,C_l\,
 {\cal K}^{(l)}_{s,\,s'}(\alpha, \alpha'),
$$
where $l$ is the index of the spherical harmonic, $ C_l = \int_0^\infty
 dz\,z^{-1}\,[J_{(l+1)/2}(z) J_{l/2}(z)]^2 $ and $J_\nu(z)$ is the
Bessel function.\footnote{For $l=1$ the coefficient $C_1$ can be
calculated analytically: $C_1=4/3\pi^2\approx 0.135$. Numerical
calculations show decreasing $C_l$ with increasing the mode number
$l$: $C_2=0.063$, $C_3=0.0373$, $C_4=0.025$, $C_5=0.018$ , and so
on.} The functions ${\cal K}^{(l)}_{s,\,s'}$ satisfy to the
condition ${\cal K}^{(l)}_{s,\,s'}(0,0)=1$; their explicit form is
given later. Then substitution of the DF in the form (\ref{eq:1.1})
leads to the following integral equation:
\begin{multline}\label{eq:1s.1}
    \phi_{s}(\alpha) =2\,\Omega_1\,\epsilon \,C_l\!\!\sum\limits_{s'=s_{\rm min}}^l
    s'^{\,2}\,D^{s'}_{l} \times \\
    \times \int\limits_0^1\dfrac{\Omega_{\rm
    pr}(\alpha')\,\alpha'\,d\,f(\alpha')/d\alpha'}
    {\omega^2-s'^{\,2}\,\Omega_{\rm pr}^2(\alpha')}\,{\cal
    K}_{\,s\,s'}^{(l)}(\alpha,\alpha')\,\phi_{\,s'}(\alpha')\,d\alpha',
\end{multline}
where $\phi_{s}(\alpha)$ is the Fourier harmonics of the radial
part of the perturbed potential (for more detail, see Paper I),
$\Omega_\textrm{pr}(\alpha)$ is the orbital precession rate,
$s_{\min}=1$ for odd $l$, and $s_{\min}=2$ for even $l$. The
coefficients $D$ are calculated by the formula
\begin{align}\label{eq:1s.1s}
    D_l^s = \left\{
    \ba{cc}
    \dfrac{1}{2^{\,2\,l}}
    \,\dfrac{(l+s)!(l-s)!}{\left[\Bigl(\frac{1}{2}\,(l-s)\Bigr)!\,
    \Bigl(\frac{1}{2}\,(l+s)\Bigr)!{\phantom{\bigg|}}\right]^2},& |l-s| \  {\rm even},\\
    \\
    0& |l-s|\ \ {\rm odd}.
    \ea
    \right.
 \end{align}

Recall that Eq. (\ref{eq:1s.1}) is written in a noninertial reference
frame centered on the mass $M_c$. Then, additional indirect potential
arising from the acceleration of the frame should be considered (see,
e.g., Tremain 2005)
$$ \Phi^i(\br,t)= {G\,\br}\int
{\br}'\frac{\delta\rho(\br',t)}{r'^3}\,dV',\ \ \delta\rho =\int \delta
f\,d\bv,
$$
where $\delta f$ is the perturbation to the background DF. Tremain
(2005) argued that for the secular perturbations, this indirect
potential must be omitted. Indeed, in studying secular evolution
one should consider perturbations $\delta f$ averaged over
Keplerian orbits. In this case the perturbed density is a
superposition of contributions of individual orbits, averaged over
their periods. A special feature of a Keplerian orbit is that the
average force from this orbit acting to the material point located
in a focus of the ellipse is equal to zero. One must be careful
however, since the perturbation is not well defined for orbits
with low angular momenta. Below we shall consider  systems with
``small amount'' of stars with low angular momenta only (see also
discussion in Sec. 2.2.1).

\bigskip

By changing the unknown function $$[\omega^2 - s^{2}\,\Omega_{\rm
pr}^2(\alpha)]\,\varphi_{s}(\alpha) = \phi_{s}(\alpha)$$ Eq.
(\ref{eq:1s.1}) can be reduced to the linear eigenvalue problem
\begin{multline}\label{eq:1s.2}
    \bigl[\,\omega^2-s^{\,2}\,\Omega_{\rm pr}^2(\alpha)\bigr]
    \varphi_{s}(\alpha) = 2\,\Omega_1\,\epsilon\,C_l  \!\sum\limits_{s'=s_{\rm min}}^l
    s'^{\,2}\,D^{s'}_{l} \times\\
    \times\int\limits_0^1 \Omega_{\rm
    pr}(\alpha')\,\alpha' \frac{d\,f(\alpha')}{d\alpha'}\,
    {\cal
    K}_{\,s\,s'}^{(l)}(\alpha,\alpha')\,\varphi_{\,s'}(\alpha')\,d\alpha'.
\end{multline}

For almost radial orbits, when $\alpha\ll 1$ or eccentricity
$e\equiv\sqrt{1-\alpha^2}\approx 1$, the precession rate is
\begin{align}\label{eq:2.5}
\Omega_{\rm pr}(\alpha) = -\frac{2\,\eps\,\Omega_1}{\pi^2}\,\alpha\,
[1+O(\alpha^2)].
\end{align}
For orbits with smaller eccentricity, the numerical coefficient
preceding the small parameter $\eps\,\Omega_1$ is somewhat greater
than $2/\pi^2$. Since one suggests that the characteristic
frequencies of the problem under consideration are of the order of
typical precession velocities, $\omega \sim \Omega_{\rm pr}\sim
\eps\Omega_1$, it is convenient to change to the dimensionless
frequencies, measured in the natural ``slow'' frequency:
\begin{align}\label{eq:1s.2s}
\bar\omega = \frac\omega{\eps\,\Omega_1},\qquad \nu(\alpha) = -\frac{\Omega_{\rm
pr}(\alpha)}{\eps\,\Omega_1}.
\end{align}
For the spherical systems, the precession is retrograde (see Tremaine 2005, or
Paper I), so $\nu(\alpha)>0$. Then the dimensionless integral equation becomes
\begin{multline}\label{eq:1s.3}
    \bigl[\,\bar\omega^2-s^{\,2}\,\nu^2(\alpha)\bigr]\,\varphi_{s}(\alpha)
    = -2\,C_l \sum\limits_{s'=s_{\rm min}}^l
    s'^{\,2}\,D^{s'}_{l} \times\\
    \times\int\limits_0^1 \nu(\alpha')\,\alpha'\,
    \frac{d\,f(\alpha')}{d\alpha'}\,
    {\cal
    K}_{\,s,\,s'}^{(l)}(\alpha,\alpha')\,\varphi_{\,s'}(\alpha')\,d\alpha',
\end{multline}

To obtain the eigenfrequency spectrum for a model it is necessary to compute
preliminarily the kernels ${\cal K}^{(l)}_{s,s'}(\alpha, \alpha')$ (universal for
all models), and the precession rate profile $\nu(\alpha)$ for the given model.
The integration over Keplerian orbits is most conveniently expressed using the
variable $\tau$, which is connected with the current radius $r$ and the true
anomaly $\zeta$ of a star\footnote{True anomaly is the angle between directions
to the star and to the pericenter.} as follows:
\begin{align}\label{eq:1s.6}
    r = {\case{1}{2}}\,R\,(1-e\cos\tau),\qquad \cos\zeta = \frac{\cos\tau
    -e}{1-e\cos\tau}.
\end{align}
Then after some transformations, the kernel ${\cal
K}^{(l)}_{s,\,s'}$ can be reduced to the form
\begin{multline}\label{eq:1s.7}
    {\cal K}^{(l)}_{s,\,s'} (\alpha, \alpha') = \frac 2
    {(2l+1)\pi^2C_l} \il_0^\pi d\tau \, r \cos(s\zeta)
     \times\\
     \times\il_0^\pi d\tau' \, r' \cos(s'\zeta')
     \,{\cal F}_l(r, r'),
\end{multline}
where $r'$ and $\zeta'$ specify the position of a star on the orbit with the
eccentricity $e'$ corresponding to the variable $\tau'$, and the notation
$$
{\cal F}_l(r, r') = \frac{\min(r,r')^l}{\max(r,r')^{l+1}}
$$
is used.

The expression for the precession rate can be obtained by
transformation of expression (4.2) of Tremaine (2005) (see also
Paper I):
\begin{align}\label{eq:1s.8}
    \nu(\alpha) = \frac \alpha{4 \pi e} \il_0^\pi \frac{\mu
    (r) (\cos\tau-e)\, d\tau}{r^2},
\end{align}
and $\nu(1)=-{\case{1}{4}}\,\pi\,\rho\,(\case{1}{2}\,R)$, where the density
$\rho(r)$ is defined by (\ref{eq:1s.10}).

For the monoenergetic models, the minimal and maximal radii are
$R_{\rm min}={\case{1}{2}}\,R\,(1-e_{\rm max}),\ \ R_{\rm
max}={\case{1}{2}}\,R\,(1+e_{\rm max}), $ where $e_{\rm
max}=(1-h^2)^{1/2}$, and $h$ is the minimal dimensionless angular
momentum corresponding to the boundary of the loss cone.

The function $\mu(r)$ is a ratio of the mass of a spherical system
inside the sphere of radius $r$ to the total mass $M_G$:
\begin{align}\label{eq:1s.9}
    \mu(r) = \frac{{\cal M}_G(r)}{M_G}, \quad {\cal M}_G(r) =
    4\pi\int\limits_{R_{\min}}^{r} r'^2 \rho(r')\,dr',
\end{align}
$M_G = {\cal M}_G(R_{\max})$, and the density is calculated by the
formula
\begin{align}\label{eq:1s.10}
    &\rho(r) = \frac{4\pi A}{r}\il_0^{L_{\max}(r)}
    \frac{f(L) \, L\,dL}{\sqrt{L^2_{\max} - L^2\phantom{\big|}}}  =
    \frac{M_G}{\pi^2 r R^2}\,\bar\rho\,(r),\nonumber\\
    &\bar\rho\,(r)=
    \il_0^{\alpha_{\max}(r)} \frac{2\alpha\,d\alpha
    \, f(\alpha)}{\sqrt{\alpha^2_{\max}-\alpha^2\phantom{\big|}}},
\end{align}
where $\alpha_{\max} = \sqrt{4\,(r/R)(1-r/R)\phantom{\big|}}$.
From here on we shall assume $R=1$.

Using (\ref{eq:1s.6}) and (\ref{eq:1s.8})\,--\,(\ref{eq:1s.10}) one can transform
the expression for the scaled precession rate $\nu(\alpha)$ to a more universal
form:
\begin{align}\label{eqv:2.12}
  \nu(\alpha)= \frac{\alpha}{2\,\pi^2 e^2}\int\limits_0^1
  d\alpha'\,\alpha'
  f(\alpha')\,{\cal Q}(\alpha,\alpha'),
\end{align}
where the kernel ${\cal Q}(\alpha,\alpha')$ doesn't depend on a DF
and equals to
\begin{align}\label{eqv:2.13}
 {\cal Q}(\alpha,\alpha')=4\int\limits_{p_{\rm min}}^{p_{\rm max}}
 dr\,\sqrt{\frac{(r-r_{\rm min})(r_{\rm max}-r)}
 {(r-r'_{\rm min})(r'_{\rm max}-r)}},
 \end{align}
 with $p_{\rm min}={\rm max}\,(r_{\rm min},r'_{\rm min}),\ \
 \ p_{\rm max}={\rm min}\,(r_{\rm max},r'_{\rm max}).
 $
Here $
 r_{\rm min}={\case{1}{2}}\,(1-e),\ \ \ r_{\rm
 max}={\case{1}{2}}\,(1+e),\ \  \
r'_{\rm min}={\case{1}{2}}\,(1-e'),\ \ \ r'_{\rm
max}={\case{1}{2}}\,(1+e'), $ and $
 e=(1-\alpha^2)^{1/2},\ \
e'=(1-\alpha'^2)^{1/2}.$ For the near radial orbits ${\cal
Q}(0,0)=4$, so that one obtains the above mentioned result
(\ref{eq:2.5}): $\nu\approx (2/\pi^2)\alpha$.

\subsection{Analytical results}

\subsubsection {Exact solution with zero frequency for the lopsided mode
($l=1$)}

Tremaine (2005) has noted that for an arbitrary distribution
$F(E,L)$ with empty loss cone, $F(E,L=0)=0$, a zero frequency
lopsided mode $l=1$ must exist. The mode corresponds to a
non-trivial perturbation arising under shift of the spherical
system as a whole relative to the central mass. The perturbed
potential in such a mode is $\delta\Phi(r,\theta)=
-\,\xi\cos\theta\,\dfrac{d\Phi_G}{dr}$, where $\xi$ is the
displacement. In terms of the function $\phi_{s=1}(\alpha)$, this
perturbation has a form
\begin{align}\label{eqv:2.15}
    \phi_1(\alpha)=\frac{e}{\alpha}\,\nu(\alpha),
\end{align}
or in terms of the function $\varphi_1(\alpha)$ from
(\ref{eq:1s.3}),
\begin{align}\label{eqv:2.16}
   \varphi_1(\alpha)=\frac{e}{\alpha\,\nu(\alpha)}.
\end{align}

One can check that (\ref{eqv:2.15}) and ${\bar\omega}=0$ provided the condition
$$f(\alpha=0)=0$$  is a solution of (\ref{eq:1s.1}) or (\ref{eq:1s.3}) for $l=1$, taking
into account the expressions (\ref{eqv:2.12}) and (\ref{eqv:2.13}), written in
the form
\begin{multline*}
 {\cal Q}(\alpha,\alpha')=-\frac{16}
 {\alpha'}\,\frac{\p}{\p\alpha'}
 \int\limits_{p_{\rm min}}^{p_{\rm max}}
 dr\,\sqrt{(r-r_{\rm min})(r_{\rm max}-r)}\times\\
 \times\sqrt{(r-r'_{\rm min})(r'_{\rm
 max}-r)},
\end{multline*}
and also the expression for the kernel ${\cal
K}_{11}^{(1)}(\alpha,\alpha')$
\begin{multline*}
 {\cal K}_{11}^{(1)}(\alpha,\alpha')=\frac{6}{e\,e'}
 \int\limits_{p_{\rm min}}^{p_{\rm max}}
 dr\,\sqrt{(r-r_{\rm min})(r_{\rm max}-r)}\times\\
 \times\sqrt{(r-r'_{\rm min})(r'_{\rm
 max}-r)}.
\end{multline*}

The lopsided solution with zero frequency is specific for
spherical systems. At the first glance, it defies common sense to
argue that the stationary mode in which the center of mass of a
spherical system does not coincide with the BH  is physical.
Indeed, it seems that movement (oscillations) of the stellar
cluster and the BH relative to the common center of mass must
occur. However, it does not occur.

It is, by all means, clear in the case of the empty  loss cone of finite
size, $h>0$ (here $h$ is a minimal value of dimensionless angular momentum
$\alpha$, for which $f(\alpha)>0$). Indeed, let us consider the spherically
symmetric cluster. Since the loss cone is finite, there is a spherical empty
cavity of finite radius in the centre of sphere. According to the Newton's first
theorem (Binney \& Tremaine 1987), in this cavity the BH does not experience a
net gravitational force from the cluster. Thus, if the BH is initially deposited
at some arbitrary point within the cavity, it would remain at this position (and
hence, acceleration of stellar cluster due to non-coincidence  of centers of mass
does not occur).

In the case when $h=0$ the situation is not so obvious, but the
net force acting to the BH from the shifted spherical system can
be zero as well. In order to assure this, one should write down
the indirect potential taking into account the expression for
perturbed density in zero lopsided mode $\delta\rho=
-\,\xi\,\cos\theta\,{d\rho}/{dr}$:
$$
\Phi^i(\br,t)=-{2\pi\,\xi\,G}\,r\,\cos\theta\int \limits_0^R dr'\,
\frac{d\rho(r')}{dr'} \int\limits_0^{\pi} \cos^2\theta'
\sin\theta'\,d\theta'=$$
$$=\frac{4\pi}{3}\,{G\,r\,\cos\theta}\,\xi\,\rho(0),
$$
Hence, the condition for omitting of the indirect potential is
$\rho(0)=0$. In what follows we suppose this condition to be
fulfilled. The conditon is not equivalent to the condition
$f(\alpha=0)=0$, imposed to the DF for the existence of such a
solution of our governing integral equation, but it is equivalent
to the stronger condition: $f(\alpha=0)=f'(\alpha=0)=0$. Indeed,
it is easy to show that if $f(\alpha)\propto\alpha^s$ for small
$\alpha$ then $\rho(r)\propto r^{(s-1)/2}$ for small $r$. So, the
condition $s>1$ must be fulfilled.

By contrast, in the disk systems the analogous $m=1$ zero mode
does not exist, because there is no analog of the Newton's first
theorem.

The very existence of zero modes is crucial for stability analysis of
spherical clusters with monotonic distributions. Indeed, the role of
destabilizing contribution of the second term in the right side of
(\ref{eq:1s.3}) falls off with increasing the number $l$. So, it is
expected that the most ``dangerous'' modes correspond to the lowest
values of $l$. But it turns out that $l=1$ mode is neutrally stable,
and the next dangerous mode $l=2$ is stable. Note, however, that such a
reasoning is not valid for systems with non-monotonic distributions.

\subsubsection{The stable mode in systems with near-radial orbits}

By analyzing (\ref{eq:1s.3}), it is easy to find one more
analytical solution with the frequency ${\bar\omega}={\cal O}(1)$
at arbitrary values of $l$, for the models with highly elongated
orbits. First of all we note that the frequency of this stable
mode corresponds to the resonance on the {\it tail} of
a narrow distribution, and so it decays exponentially
slowly. In this way the mode differs from the unstable modes of
interest which have a resonance in a region where the distribution
is localized, i.e. at $\alpha\la \alpha_T$; so they have
characteristic frequencies and growth rates of the order of ${\cal
O}(\alpha_T)$.

After setting ${\bar\omega} \sim 1 \gg \alpha_T$ in (\ref{eq:1s.3}), omitting the
second summand in l.h.s., turning to the spoke approximation, and taking into
account the equality $\sum\limits_{s\,=\,s_{\rm min}}^l s^2 D_l^s
={\case{1}{4}}\,l\,(l+1) $, one finds
\begin{align}\label{eqv:3.4}
{\bar\omega}^2= \frac{2\,C_l}{\pi^2}\, l\,(l+1).
\end{align}
It is essential that this {\it high-frequency} mode is independent
of details of the DF. Note also that in the systems with prograde
precession, this mode describes the well-known radial orbit
instability (instead of the neutral oscillations).

\subsubsection{The Variational principle}

Using (\ref{eq:1s.3}), one can prove two important statements:
\begin{enumerate}
    \item For spherical system models with monotonic distributions
$f(\alpha)$, the eigenfrequency squared, $\bar \omega^2$, must be
a real number. This means that either the instability is absent at
all, or aperiodic instability with $\textrm{Re}\,\bar\omega=0$
occurs.
    \item Rotating (or oscillating) unstable modes may appear only in models
with non-monotonic distributions.
\end{enumerate}

Let us write Eq. (\ref{eq:1s.3}) in the form
\begin{multline}\label{eq:4.1}
    \bar\omega^2 \varphi_{s}(\alpha) =
    s^{\,2}\,\nu^2(\alpha) \varphi_{s}(\alpha) - 2\,C_l\!\!\!\!\sum\limits_{s'=s_{\rm min}}^l
    s'^{\,2}\,D^{s'}_{l} \times\\
    \times \int\limits_0^1 g(\alpha')\,{\cal
    K}_{\,s\,s'}^{(l)}(\alpha,\alpha')\,\varphi_{\,s'}(\alpha')\,d\alpha',
\end{multline}
where $ g(\alpha) = \nu(\alpha)\,\alpha\,\,{df(\alpha)}/{d\alpha}. $ We multiply
both parts of Eq. (\ref{eq:4.1}) by $s^2 D_l^s\varphi_s^{\ast}(\alpha)$, sum the
result over $s$ (asterisk means the complex conjugation), and integrate over
$\alpha$ with the weight $g(\alpha)$. Then we obtain
\begin{multline}\label{eq:4.2}
{\bar\omega}^2\int\limits_0^1 g(\alpha)\,d\alpha\sum\limits_{s=s_{\rm min}}^l
s^2D_l^s\,|\varphi_s(\alpha)|^2\, =\\
=\int\limits_0^1\nu^2(\alpha)\,g(\alpha)\, d\alpha\sum\limits_{s=s_{\rm min}}^l
s^4
D_l^s\,|\varphi_s(\alpha)|^2\,-\\
- 2C_l \int\limits_0^1 d\alpha\int\limits_0^1 d\alpha' \sum\limits_{s=s_{\rm
min}}^l \sum\limits_{s'=s_{\rm min}}^l (ss')^2 D_l^s\,
D_l^{s'}\times\\
\times{\cal
K}^{(l)}_{s\,s'}(\alpha,\alpha')\,[g(\alpha)\varphi_s^{\ast}(\alpha)]
\,[g(\alpha')\,\varphi_{s'}(\alpha')].
\end{multline}
The reality of the coefficients of ${\bar\omega^2}$ in the l.h.s.
of (\ref{eq:4.2}) and the first term in the r.h.s. is evident.
With the help of (\ref{eq:1s.7}), one can show that the kernel in
Eq. (\ref{eq:4.2}) has the following property of symmetry:
\begin{align}\label{eq:4.3}
{\cal K}_{\,s,\,s'}^{(l)}(\alpha,\alpha')={\cal
K}_{\,s',\,s}^{(l)}(\alpha',\alpha).
\end{align}
So, one can readily see that the second term in the r.h.s is real
also. Consequently, taking the imaginary part of Eq.
(\ref{eq:4.2}), one obtains
 \begin{align}\label{eq:4.4}
 {\rm Im}({\bar\omega}^2) \int\limits_0^1 g(\alpha)\,d\alpha\sum\limits_{s=s_{\rm min}}^l
s^2D_l^s\,|\varphi_s(\alpha)|^2\equiv 0.
 \end{align}

>From the last equality, the statements formulated above follow immediately. If
the function $g(\alpha)$ (or, equivalently, $df(\alpha)/d\alpha$) is
constant-sign, then the integral should be non-zero, and so ${\rm
Im}({\bar\omega}^2)=0$. In contrast, when ${\rm Im}({\bar\omega}^2) \ne 0$, the
integral must be equal to zero. Consequently, the function $g(\alpha)$ should
change its sign, i.e. DF $f(\alpha)$ is non-monotonic.

Let us explain the term {\it variational principle} used in the
title of this subsection. Consider a dynamic equation in the form
$d^2\xi/dt^2\equiv -\omega^2\xi=-{\hat K}\xi$. Provided that
``elasticity operator'' ${\hat K}$ is Hermitian, the dynamic
equation may be obtained from the conditions
$\delta(\omega^2)/\delta\xi=0$ and
$\delta(\omega^2)/\delta\xi^{\ast}=0$. Here $\delta\xi$ and
$\delta\xi^{\ast}$ are considered formally as independent
variations while the functional $\omega^2$ is
$$
\omega^2=\frac{\int\xi^{\ast}\!\!\,({\hat K}\xi)\,
w(\alpha)\,d\alpha}{\int |\xi|^2 w(\alpha)\,d\alpha}
$$
($w(\alpha)$ is a nonnegative weight function). In such a case it is used to
speak about the variational (or energy) principle (see, e.g., review by Kadomtsev
1966 on MHD-stability of plasma). From the other hand it is easy to see
that if $\int |\xi|^2 w(\alpha)\,d\alpha \ne 0$ for any nontrivial $\xi$ then
$\omega^2$ is real. Thus usually (as is the case in MHD-stability theory of plasma
where  ${\hat K}$ is Hermitian and  $w>0$) the notions ``variational principle''
and reality of $\omega^2$ are identical. However, in our case the condition $\int
|\xi|^2 w(\alpha)\,d\alpha\ne 0$ is not satisfied for any nontrivial $\xi$ unless
the DF is monotonic. Assuming that this condition is fulfilled, following the
tradition that originates from plasma physics we speak that the variational
principle takes place. Only in this case the dynamical equation can be
interpreted mechanically, in terms of elastic forces.

Evidently, the condition (\ref{eq:4.4}) is a serious obstacle to obtain
unstable rotating modes. So, one might want to get round this obstacle.
For instance, if we slightly change the initial monotonically
increasing DF in a narrow region near $\alpha=1$, to vanish quickly but
smoothly, then a modified system would be practically indistinguishable
from the initial one. But then the variational principle breaks down.
The question can be formulated as follows: May the discontinuous
vanishing of $f(\alpha)$ at $\alpha=1$ be considered as the violation of
monotony for the DF?

Importance of this question is known since stability study of stellar systems
with isotropic DFs, $F=F(E)$ (Antonov, 1960, 1962). The variational principle
there required a DF to be decreasing function of energy $E$, $F'(E)<0$, everywhere.
The systems with $F'(E)>0$ need separate examinations, that was carried out in
some cases (see, e.g., Antonov, 1971, Kalnajs, 1972, Polyachenko and Shukhman,
1972, 1973, Fridman and Polyachenko, 1984). An essential difference of the second
type of DFs is in jumps to zero at the phase space boundary $E=E_{\rm bound}$. In
fact, we have in this case an interval degenerated into the single point
$E=E_{\rm bound}$ where $F(E)$ is decreasing.

We checked numerically a possibility of the instability development connected
with the maximum on the edge of the distribution function's domain. For this
purpose, number of models smoothed near $\alpha=1$  were computed.
The computations showed no sign of instability, in contrast to
isotropic distributions, $F=F(E)$. The reason for the tolerance of
our present models is in fact that the kernels ${\cal K}$ of
integral operators in (\ref{eq:4.1}) vanish for the circular
orbit $\alpha=1$, so details of the initial distribution near
circular orbits cannot affect much solutions of the integral
equation (\ref{eq:4.1}).

Roles of different terms in Eq. (\ref{eq:4.1}) can be easily
understood. When $\p F/\p L>0$, the first term of the right side in
Eq. (\ref{eq:4.1}) favors stabilization, while the second term
gives destabilization (taking into account that the operator
involved into this construction is self-adjoint and positively
defined). In principle, this destabilizing effect could lead to
instability. However, this is not the case because the stabilizing
contribution exceeds destabilizing one in all cases considered
both by Tremaine (2005), and in the present paper (see the
following sections).

\subsection{Unstable models}

Instability boundaries in terms of the angular momentum dispersion
$\alpha_T<(\alpha_T)_c$ found in Paper~I for the monoenergetic DF with
\begin{align}\label{eq:2.1}
f(\alpha) = \frac{N}{\alpha_T^2} \(\frac{\alpha^2}{\alpha_T^2}\)^n
\exp(-\alpha^2/\alpha_T^2),
\end{align}
($N$ is the normalization constant, $\alpha_T$ is the
dimensionless angular momentum dispersion, $n$ is the real number)
have a qualitative character only: formally, these boundaries lie
outside the validity of the spoke approximation, since
$(\alpha_T)_c \sim 1$. Obtaining such critical dispersions means
only that the spoke models, in which $\alpha_T\ll 1$ by
definition, are certainly unstable. So the quantitative
determination of these boundaries with help of the exact integral
equation is required. The {\it power\,--\,exp} model
(\ref{eq:2.1}) is studied in Sec. 2.3.1.

In Sec. 2.3.2  we study a simplest {\it Heaviside} model
consisting of two steps (at $\alpha=h_1$ and at $\alpha=h_2$)
(both in the spoke approximation framework and using exact
integral equation):
\begin{align}\label{eqv:2.2}
f(\alpha)=\frac{2}{h_2^2-h_1^2}\,\,\Bigl[H(\alpha-h_1)-H(\alpha-h_2)\Bigr],\
\ h_1<h_2<1
\end{align}
($H(\alpha)$ denotes the Heaviside function). Finally, in Sec.
2.3.3 we consider the {\it log\,--\,exp} model with DF
\begin{align}\label{eq:2.3}
f(\alpha) = \frac{N}{\alpha_T^2} \ln(\alpha^2/h^2) \exp(-\alpha^2/\alpha_T^2),
\end{align}
for $\alpha \geq h$, and $f(\alpha)=0$ for $\alpha < h$, i.e.,
with the empty loss cone ($N$ is the normalization constant).

\subsubsection{The power\,--\,exp model}

Following Paper I, here we consider the stability of models
(\ref{eq:2.1}) with $n=2$ and $n=3$ relative to the spherical
harmonic $l=3$. We remind that at the limit $\alpha_T\ll 1$,  both
these models were unstable (the stability boundaries obtained
using spoke approximation were $(\alpha_T)_c = 0.193$ for $n=2$,
and $(\alpha_T)_c = 0.283$ for $n=3$).

For distribution (\ref{eq:2.1}) one finds
$$
N^{-1} = \frac12 \il_0^{1/\alpha_T^2} z^n\exp(-z)\, dz,\ \ \ z\equiv
\frac{\alpha^2}{\alpha_T^2}.
$$
Particularly, in the case $\alpha_T\ll 1$, the normalization constant is $N
=2/(n\,!)$\,. From (\ref{eq:1s.10}), we obtain
$$
\bar \rho\,(r) = N \il_0^{\alpha^2_{\max}(r)/\alpha_T^2} \frac{z^n
e^{-z}\, dz}{\sqrt{\alpha^2_{\max}-z\,\alpha^2_T}} .
$$
Further calculations of the density (\ref{eq:1s.10}) and
precession rate (\ref{eqv:2.12}) profiles should be evaluated
numerically.

\begin{figure}
 \centerline{\includegraphics[width=102mm, draft=false]{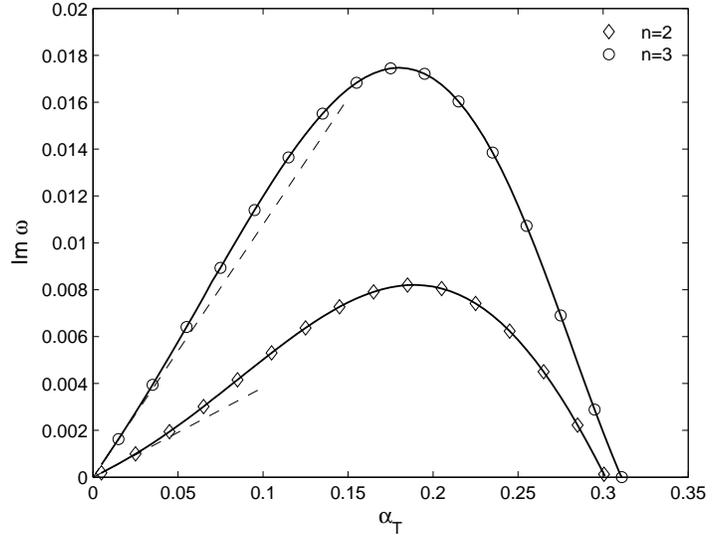}}
 \caption{The dependence of the growth rate ${\rm Im}\,(\bar\omega)$ vs.
dimensionless angular momentum dispersion $\alpha_T$ of the mode
$l=3$ for models $n=2$ (diamonds) and $n=3$ (circles). Dashed
lines show the asymptotic behavior obtained using spoke
approximation equation (\ref{eq:2.power_sa}):
$\Im({\bar\omega}/\alpha_T)=(2/\pi^2)\,\Im\lambda=0.189$ and 0.532 for
$n=2$ and 3 respectively (exact solution for $\alpha_T=0.003$
gives 0.185 and 0.529).}
 \label{fig:0}
\end{figure}

Solutions of the integral equation (\ref{eq:1s.3}) for $n=2$ and $n=3$ show
similar behavior. At small values of $\alpha_T$, each model has one unstable
mode. With increasing the dimensionless angular momentum dispersion $\alpha_T$,
the growth rate of the instability decreases, and then vanishes at some critical
value $(\alpha_T)_c$: for the model $n=2$, $(\alpha_T)^{(2)}_c \simeq 0.301$, for
the model $n=3$, $(\alpha_T)^{(3)}_c \simeq 0.311$ (see Fig.\,\ref{fig:0}). We
conclude that the spoke approximation in this case is qualitatively correct, but
quantitatively poor. The instability becomes saturated at some critical value
$(\alpha_T)_c$, while the discrepancy between exact and
approximate values of $(\alpha_T)_c$ for both models are not small.

Apart from the unstable mode, the spectrum of each model has a discrete mode, the
growth rate of which is equal to zero at small $\alpha_T$, and becomes
negative with increasing $\alpha_T$. This is just that weakly decaying mode with
the frequency $ {\bar\omega}^2 \approx 2\,C_l\,l\,(l+1)/\pi^2$ (at $\alpha_T \ll
1$) which was mentioned in Sec. 2.2.2. The dependence of the frequency on $l$ for
this mode was one of the tests for numerical code of the integral equation solver.
Another test was detecting the zero lopsided mode $l=1$ mentioned in
Sec. 2.2.1.

The third test was the evaluation of $\bar\omega\,(\alpha_T)$ dependence in the
spoke approximation limit. Assuming that $\bar\omega = 2\lambda\alpha_T/\pi^2$,
and using ${\cal K}_{s,s'}^{(l)}(\alpha,\alpha') \approx 1$,
$\phi_s(\alpha)\approx 1$, and $\nu(\alpha)\approx 2\alpha/\pi^2$ in
(\ref{eq:1s.1}), one can obtain the equation for the $l=3$ mode
\begin{align}\label{eq:2.power_sa}
    \il_0^\infty \rmn{d}z \, (n-z) z^n e^{-z} \(\frac1{\lambda^2-z} +
    \frac{15}{\lambda^2-9z}\) = {\cal O}(\alpha_T^2).
\end{align}
By setting the r.h.s. to zero, one obtains an unstable mode for
each $n$: $\lambda = 2.243 + 0.189i$ for $n=2$, and $\lambda =
2.592+0.532i$ for $n=3$. The same values obtained from solution of
the exact integral equation (\ref{eq:1s.3}) for $\alpha_T=0.003$
are $\lambda = 2.240+0.185i$ and $\lambda = 2.588+0.529i$,
correspondingly.

\subsubsection{The Heaviside model}

The simplest non-monotonic model that allows to progress rather
far by analytical methods is the model with a piecewise constant
distribution over momentum (\ref{eqv:2.2}). In other words, we
assume the DF to be non-zero only within the interval
$h_1<\alpha<h_2$, where it is taken constant.

When studying stability of discontinuous distributions such as
(\ref{eqv:2.2}), it is more convenient to start with the integral
equation in the form (\ref{eq:1s.1}). Substituting (\ref{eqv:2.2})
into Eq. (\ref{eq:1s.1}), one obtains
\begin{multline}\label{eq:2.15}
\phi_{s}(\alpha) = \frac{4\,C_l\,\eps\,\Omega_1}{h_2^2-h_1^2}
  \sum\limits_{s'=s_{\rm min}}^l\!\!\! s'^{\,2}\,D_{l}^{s'}
\left[\dfrac{\Omega_{\rm pr}(h_1)\,h_1}
{\omega^2-s'^{\,2}\,\Omega_{\rm pr}^2(h_1)}\right.\times\\
\times {\cal K}_{\,s\,s'}^{(l)}(\alpha,h_1)\, \phi_{\,s'}(h_1)
-\dfrac{\Omega_{\rm pr}(h_2)\,h_2}
{\omega^2-s'^{\,2}\,\Omega_{\rm pr}^2(h_2)}\times\\
\times\left.{\cal K}_{\,s\,s'}^{(l)}(\alpha,h_2)
\,\phi_{\,s'}(h_2)\right].
\end{multline}
Let us turn again to the natural slow scale of frequencies
according (\ref{eq:1s.2s}) and then substitute in (\ref{eq:2.15})
particular values $\alpha=h_1$ and $\alpha=h_2$. For brevity sake,
the following designations are used: $\nu_1 \equiv \nu(h_1)$, $\nu_2
\equiv \nu(h_2)$. We have
\begin{multline}\label{eq:2.17}
\phi_{s}(h_1) =-\frac{4\,C_l}{h_2^2-h_1^2}\sum\limits_{s'=s_{\rm min}}^l\!\!\!
s'^{\,2}\,D_{l}^{s'} \left[\dfrac{\nu_1\,h_1} {\bar
\omega^2-s'^{\,2}\,\nu_1^2}\,{\cal K}_{\,s\,s'}^{(l)}(h_1,h_1)\right.\times\\
\times\left.\phi_{\,s'}(h_1)- \dfrac{\nu_2\,h_2}
{\bar\omega^2-s'^{\,2}\,\nu_2^2}\,{\cal
K}_{\,s\,s'}^{(l)}(h_1,h_2)\,\phi_{\,s'}(h_2)\right],
\end{multline}
\begin{multline}\label{eq:2.18}
\phi_{s}(h_2) =-\frac{4\,C_l}{h_2^2-h_1^2}\sum\limits_{s'=s_{\rm min}}^l\!\!\!
s'^{\,2}\,D_{l}^{s'} \left[\dfrac{\nu_1\,h_1}
{\bar\omega^2-s'^{\,2}\,\nu_1^2}\,{\cal
K}_{\,s\,s'}^{(l)}(h_2,h_1)\right.\times\\
\times\left.\phi_{\,s'}(h_1)- \dfrac{\nu_2\,h_2}
{\bar\omega^2-s'^{\,2}\,\nu_2^2}\,{\cal
K}_{\,s\,s'}^{(l)}(h_2,h_2)\,\phi_{\,s'}(h_2)\right].
\end{multline}
This set of equations relative to $\phi_s(h_1)$ and $\phi_s(h_2)$,
($s=1,2,...,[\frac{1}{2}\,(l+1)]$) can be reduced to the standard
linear set. Introducing new unknown functions
\[X_s=\frac{\nu_1\,h_1}{\bar\omega^2-s^2\nu_1^2}\,\,\phi_s(h_1),\ \ \
Y_s=\frac{\nu_2\, h_2}{\bar\omega^2-s^2\nu_2^2}\,\,\phi_s(h_2),
\]
one obtains
\begin{multline}\label{eq:2.20}
 \left(\bar\omega^2-s^2\nu_1^2\right)\,X_s=-4\,C_l\,
\frac{\nu_1\,h_1}{h_2^2-h_1^2} \sum\limits_{s'=s_{\rm
min}}^l\!\!\! s'^{\,2}\,D_{l}^{s'} \times\\
\times\left[{\cal K}_{\,s\,s'}^{(l)}(h_1,h_1)\,X_{s'} - {\cal
K}_{\,s\,s'}^{(l)}(h_1,h_2)\,Y_{s'}\right],
\end{multline}

\begin{multline}\label{eq:2.21}
\left(\bar\omega^2-s^2\nu_2^2\right)\,Y_s=-4\,C_l\,
\frac{\nu_2\,h_2}{h_2^2-h_1^2} \sum\limits_{s'=s_{\rm
min}}^l\!\!\! s'^{\,2}\,D_{l}^{s'} \times\\
\times\left[{\cal K}_{\,s\,s'}^{(l)}(h_2,h_1)\,X_{s'} - {\cal
K}_{\,s\,s'}^{(l)}(h_2,h_2)\,Y_{s'}\right].
\end{multline}

The precession rates in these equations can be expressed through the complete
elliptical integrals $\textbf{K}$ and $\textbf{E}$:
 \[
 \nu_1=2\,C_1\,\frac{h_1}{e_1}\,\frac{1-q^2\,Q(q)}{1-q^2}, \qquad
 \nu_2=2\,C_1\,\frac{h_2}{e_1}\,\frac{Q(q)-q}{1-q^2},
 \]
where
 $C_1={4}/(3\pi^2),\ \ q={e_2}/{e_1},\ \ e_1=(1-h_1^2)^{1/2}, \ \
 e_2=(1-h_2^2)^{1/2},$ and the function $Q(q)$ is
 \[Q(q)=\frac{1}{2\,q^2}\,\left[\,(1+q^2)\,{\bf E}\,(q)-(1-q^2)\,{\bf K}\,(q)\right].
  \]
In the limit $h_2\to h_1$, the frequencies $\nu_1$ and $\nu_2$ are
coincident: $ \nu_1=\nu_2=({2}/{\pi^2})\,({h}/{e})$, where
$h=h_1=h_2$, and $e=e_1=e_2$. Note that a determinant of the set
of equations (\ref{eq:2.20}) and (\ref{eq:2.21}) has a rank
$2\,[\frac{1}{2}\,(l+1)]$. Particularly, for the mode $l=1$, the
rank is equal to 2. Roots of the determinant are calculated
numerically. The difference $h_2-h_1$ has a meaning of dispersion,
i.e., it is analogous to the parameter $\alpha_T$ in our models
with smooth distributions.

A simple analytical task is to ascertain that ${\bar\omega}^2=0$
is the eigenvalue of the determinant for $l=1$.  We have for
${\cal K}_{11}^{(1)}(\alpha,\alpha')$
\begin{align}\label{eq:2.27}
    {\cal K}_{11}^{(1)}(\alpha,\alpha')=e_<\,Q(\kappa),
\end{align}
where $ \kappa={e_<}/{e_>}, \ \ e_<={\rm min}\,(e,e'),\ \ e_>={\rm
max}\,(e,e'), \ \ e=(1-\alpha^2)^{1/2},\ \
e'=(1-\alpha'^2)^{1/2}$. In particular,
\begin{align}\label{eq:2.29}
{\cal K}_{11}^{(1)}(h_1,h_1)=e_1,\ \ {\cal
K}_{11}^{(1)}(h_2,h_2)=e_2,
\end{align}
\begin{align}\label{eq:2.30}
{\cal K}_{11}^{(1)}(h_1,h_2)={\cal
K}_{11}^{(1)}(h_2,h_1)=e_2\,Q(q),\ \ q={e_2}/{e_1}.
\end{align}
Setting ${\bar\omega}^2=0$ in the determinant of the set
(\ref{eq:2.20}) and (\ref{eq:2.21}), and using the expressions for
the elements of the kernel (\ref{eq:2.29}) and (\ref{eq:2.30}), we
can show that it is equal to zero identically. This just means the
occurrence of a zero mode in the spectrum. Another root
${\bar\omega}^2$ for $l=1$ is positive for any values of $h_1$ and
$h_2$, which agrees with our previous conclusion (Paper I) that
the instability is absent for the mode $l=1$.

\begin{figure}
 \centerline{\includegraphics[width=102mm, draft=false]{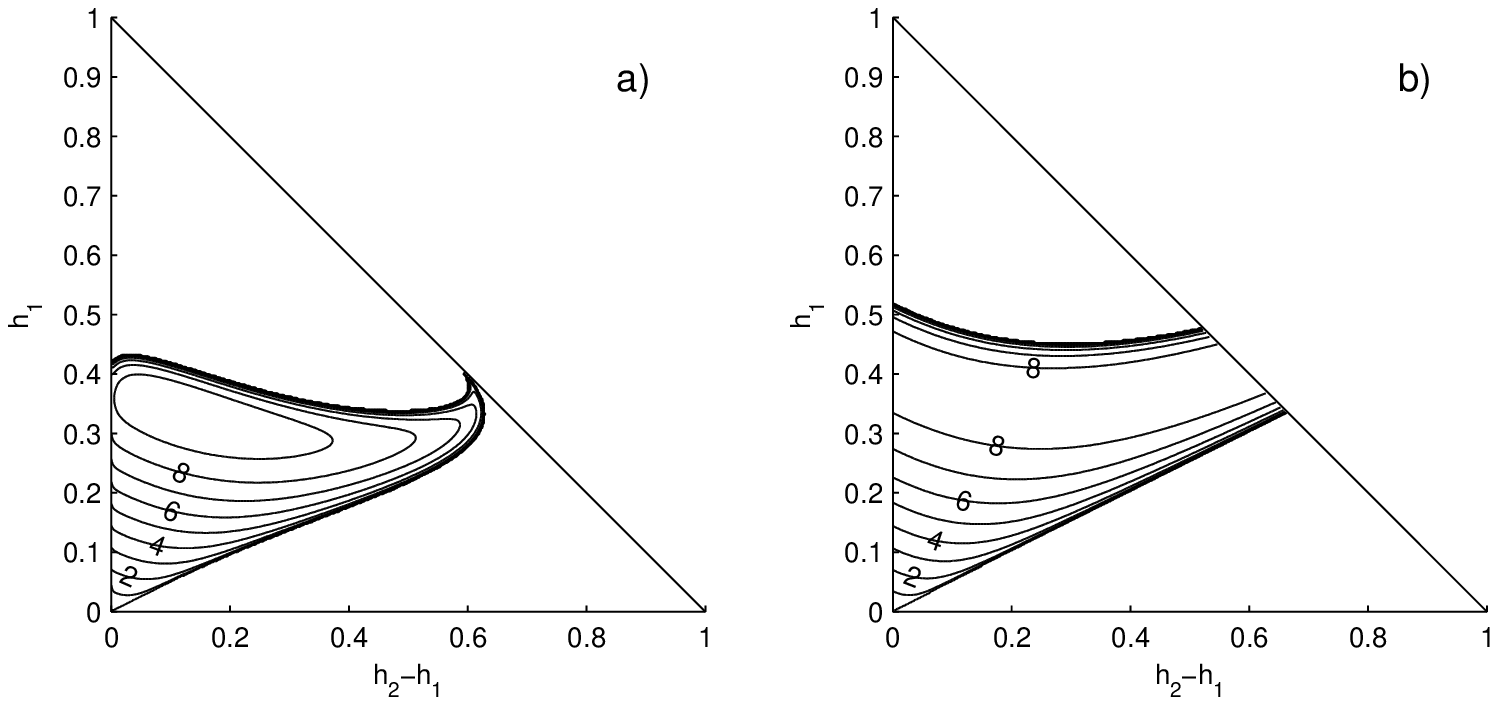}}
 \vspace{-1mm}
 \centerline{\includegraphics[width=102mm, draft=false]{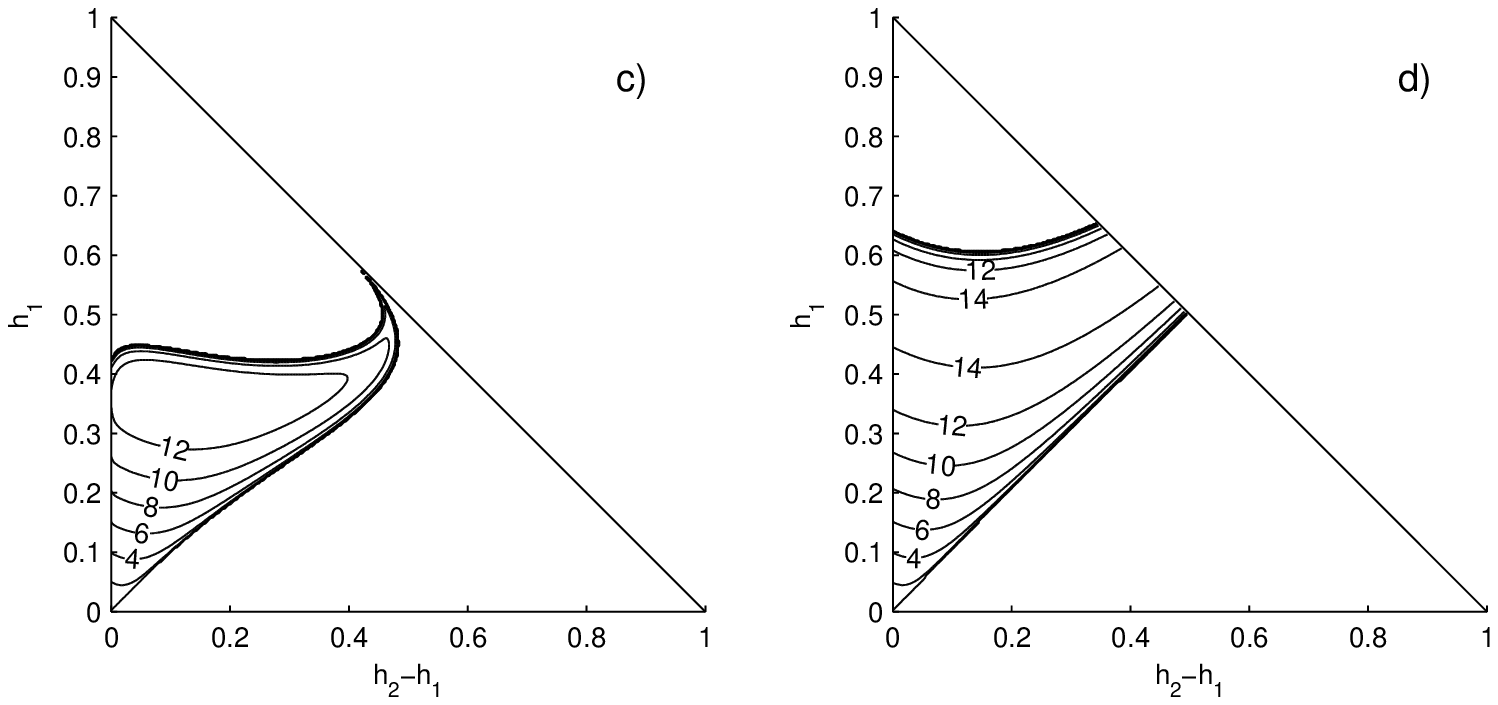}}
 \vspace{-1mm}
 \centerline{\includegraphics[width=102mm, draft=false]{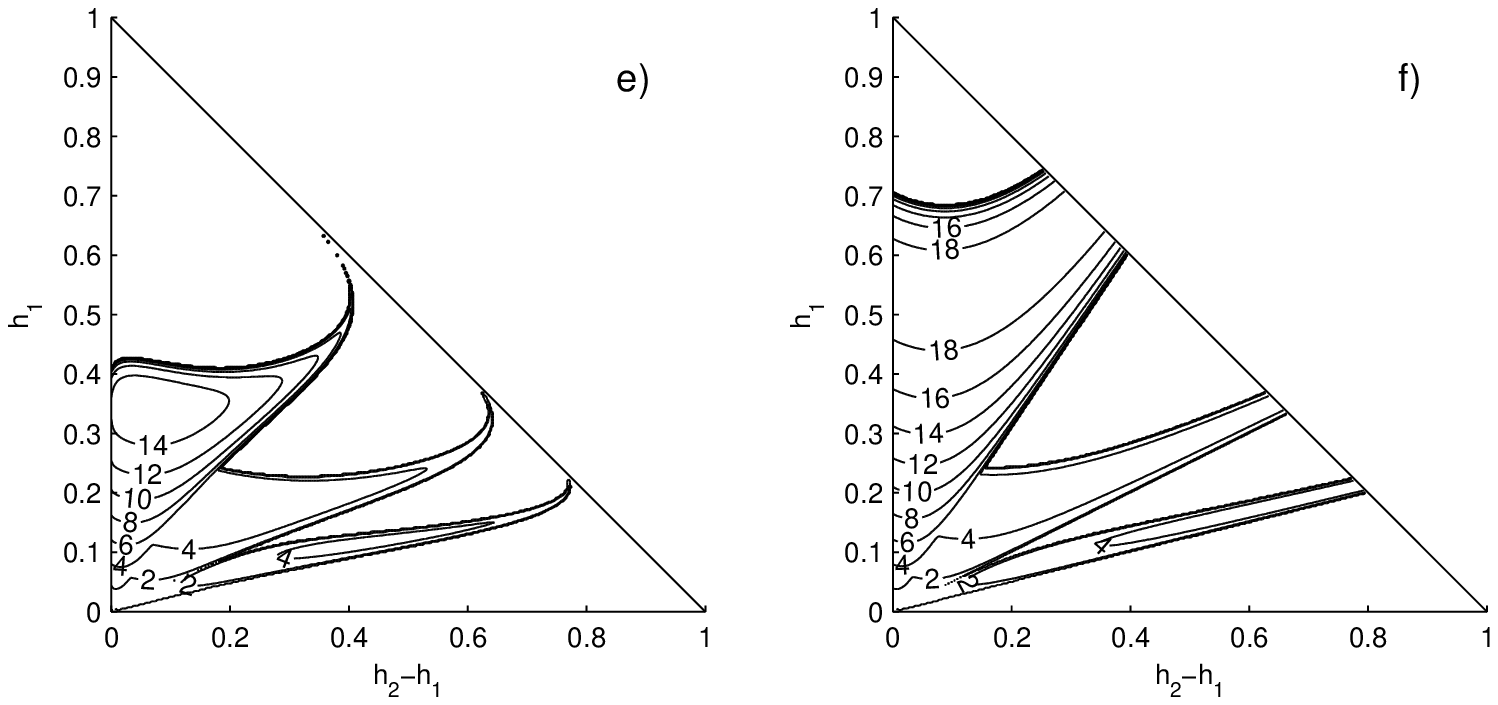}}
 \caption{The stability boundaries and the isolines
$10^2\,{\rm Im}\,({\bar\omega})$ on the plane $(h_2-h_1, h_1)$ for the Heaviside
model (left: exact calculations, right: spoke approximation calculations): {\it
a} and {\it b} -- for the mode $l=3$; {\it c} and {\it d} -- for the mode $l=4$,
{\it e} and {\it f} -- for mode $l=5$.  The spoke approximation is reliable in
the lower left corner of the domain. It is also seen that for $l=3$, the growth
rate sharply decreases when the ratio of the difference $h_2 - h_1$ (the
analog of the dispersion $\alpha_T$ for models with smooth DFs) to the size of
the loss cone, $h_1$, becomes greater than 2.2.}
 \label{fig:1}
\end{figure}

It is useful to derive equations in (\ref{eq:2.20}) and
(\ref{eq:2.21}) in the spoke limit, when the distribution is
localized in a region of small $\alpha$. This means that
we suggest $h_1\ll 1$, $h_2\ll 1$, $h_1<h_2$, set in
(\ref{eq:2.15}) $\phi_s(\alpha)=(-1)^s$, ${\cal
K}_{s,\,s'}^{(l)}=(-1)^{s+s'}$, and finally obtain
\begin{align}\label{eq:2.31}
1 =-\frac{4\,C_l}{h_2^2-h_1^2}\sum\limits_{s'=s_{\rm min}}^l\!\!\!
s'^{\,2}\,D_{l}^{s'} \left[\dfrac{\nu_1\,h_1} {\bar\omega^2-s'^{\,2}\,\nu_1^2}-
\dfrac{\nu_2\,h_2} {\bar\omega^2-s'^{\,2}\,\nu_2^2}\,\right].
\end{align}
 For
the precession frequencies $\nu_1$ and $\nu_2$, one has in this limit
$\nu_1=({2}/{\pi^2})\,h_1$, $\nu_2=({2}/{\pi^2})\,h_2.$ Introducing
${\tilde\omega}=\ccase{1}{2}\,\pi^2\,{\bar\omega}$, let us write down, e.g., the
characteristic equation for the mode $l=3$. In this case $D_3^1=\frac{3}{16}$, $\
D_3^3=\frac{5}{16}$, hence
\begin{multline}\label{eq:2.34}
1=-\frac{3\pi^2 C_3}{8}\,\frac{1}{h_2^2-h_1^2}
\left[\frac{h_1^2}{{\tilde\omega}^2-h_1^2}+\frac{15
h_1^2}{{\tilde\omega}^2-9h_1^2}-\right.\\
- \left.\frac{h_2^2}{{\tilde\omega}^2-h_2^2}-\frac{15
h_2^2}{{\tilde\omega}^2-9h_2^2}\right].
\end{multline}
Due to the denominator $h_2^2-h_1^2\ll 1$, the role of ``self-gravity'' may be
made sufficiently large in spite of small parameter $C_3$. This may give the
oscillating instability under certain conditions for $h_1$ and $h_2$. The
limiting solutions serves a test for the model with arbitrary $h_1$ and $h_2$.
\bigskip

The results for the modes $l=3$, $l=4$ and $l=5$ are presented in
Fig.\,\ref{fig:1}{\it a\,--\,f}. They show the boundaries of
instability domains on the plane $(h_2-h_1, h_1)$. Left panels
show the results of computations from the exact set of equations
(\ref{eq:2.20}) and (\ref{eq:2.21}); right panels -- the results
from the spoke equation for this model (\ref{eq:2.31}). It is seen
that in the region $h_1\ll 1$, $h_2-h_1\ll 1$, the results
obtained from the spoke equation and those from the exact
equations do coincide. Location of the growth rate maxima at
Figs.\,\ref{fig:1}{\it a}, 1{\it c}, 1{\it e}, as well as the
values of growth rates are practically the same.\footnote{Two
additional instability domains for $l=5$ are explained by more
complicated structure of the characteristic equation for this
mode, compared to the modes $l=3$ and $l=4$.} We conclude that for
the non-monotonic DF the instability is insensitive to a number
$l$ of the mode. This is a characteristic feature for the
loss-cone instability.  Recall that in models with monotonic
distributions, the destabilizing term quickly decreases with
increasing spherical number $l$ of the mode.

\subsubsection{The log\,--\,exp model}

In some numerical models (see, e.g., Cohn, Kulsrud, 1978, Berczik
et. al. 2005, Spurzem et. al. 2005) the initially isotropic distribution transforms
under action of a massive black hole into one monotonically
increasing with angular momentum, $f(\alpha)\propto
\ln(\alpha/h)$, where $h$ defines the minimum angular momentum of
a star which is not absorbed by the black hole. In this section,
we consider the stability of DF (\ref{eq:2.3}). For it, one finds
\begin{multline*}
 N^{-1} = \il_h^1 \frac{d\alpha\, \alpha}{\alpha_T^2} \ln
\(\frac{\alpha^2}{h^2}\) e^{-\alpha^2/\alpha^2_T} = \frac12
\left[-\textrm{Ei}(-h^2/\alpha_T^2)+\right.\\
  +
\textrm{Ei}(-1/\alpha_T^2) + \left. \ln
(h^2)\,e^{-1/\alpha_T^2}\right],
\end{multline*}
where $\textrm{Ei}(z)$ is the exponential integral. Density
profile is obtained from (\ref{eq:1s.10}):
$$
{\bar\rho}(r) = \frac{N}{\alpha_T^2}
\il_{h^2}^{\alpha^2_{\max}(r)} \frac{dz}{\sqrt{\alpha^2_{\max}(r)
- z\phantom{\big|}}}\,\ln \(\frac z{h^2}\) \, e^{-z/\alpha_T^2}.
$$
Much as in the power\,--\,exp model considered above, the
calculations of the precession rate should be performed
numerically.

The qualitative pattern of the spectrum for this model is similar
to that of the power\,--\,exp model: when the dispersion is not
too large, two discrete modes occur, one of which being unstable.

\begin{figure}
 \centerline{\includegraphics[width=45mm, draft=false]{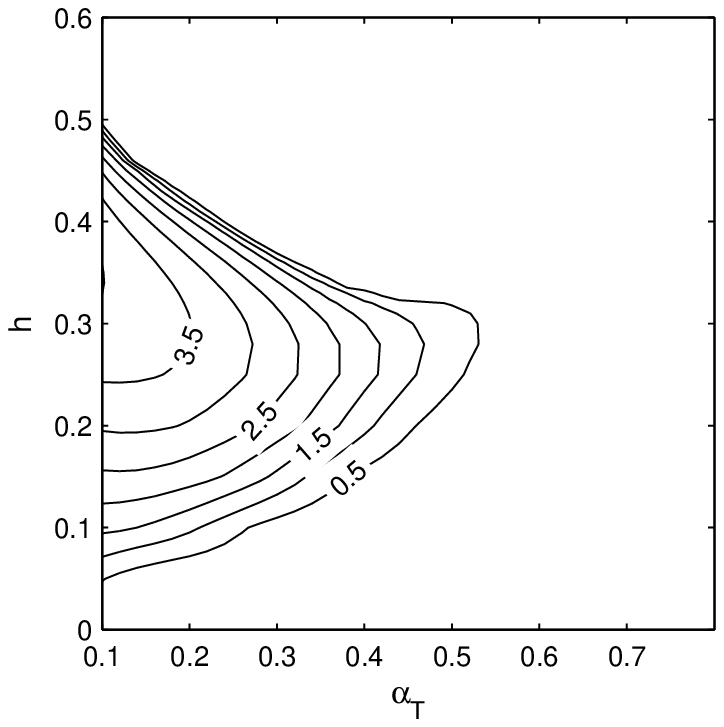}
 \includegraphics[width=45mm, draft=false]{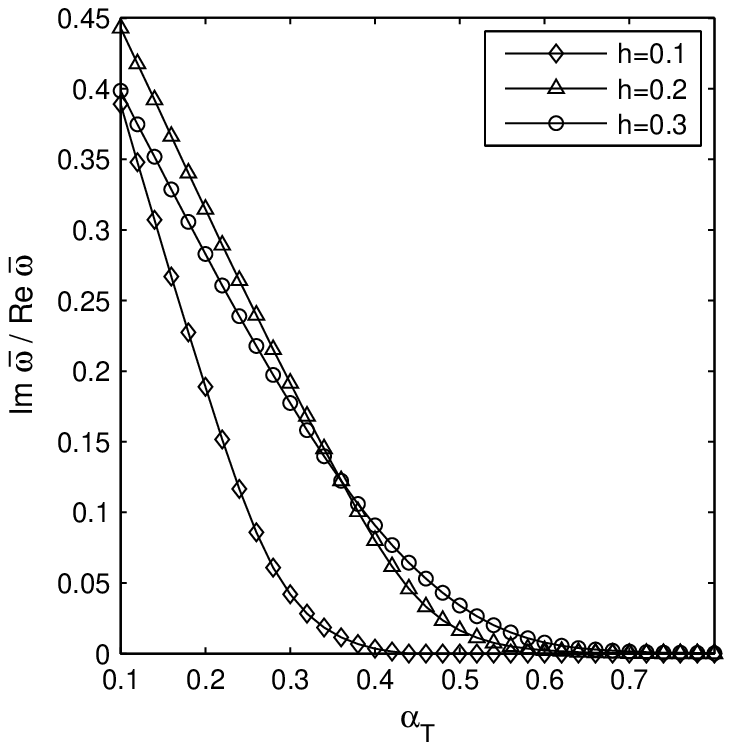}}
\caption{The mode $l=3$ for the log\,--\,exp model. Left: isolines
$10^2\,\textrm{Im}\,\bar\omega$ on the plane $(\alpha_T, h)$. Right: the ratio of
the growth rate $\textrm{Im}\,(\bar\omega)$ to the real part of the frequency
$\textrm{Re}\,\bar\omega$, for values $h=0.1,\, 0.2,\, 0.3$.}
 \label{fig:2}
\end{figure}

In Fig.\,\ref{fig:2} (left panel), the isolines
$10^2\,\textrm{Im}\,{\bar\omega}$ on the plane of parameters
$(\alpha, h)$ for the mode $l=3$ are presented. Comparison of the figures \ref{fig:1} and \ref{fig:2} shows
qualitative coincidence of growth rates behavior on the dispersion of DF and the  size of the loss cone in this and Heaviside models.
The right panel shows the ratio of imaginary part to real part of
$\bar \omega$ vs. dimensionless angular momentum dispersion
$\alpha_T$ for several values of loss cone size parameter $h$.

\subsection{Stable models}

Instability of spherical clusters around massive black holes was first studied by Tremaine (2005). He considered distributions of the form
\begin{align}\label{eq:3.1}
f(I_1,I_2) \propto I_1^{\,b}\, \ln \(\frac{I_2}{h\,I_1}\)
\end{align}
in the domain $I_{\min} \le I_1\le I_{\max}$, $I_2>hI_1$ (and zero
outside this domain); $I_r$, $I_2=L$ are the action variables,
$I_1=I_r+L$; $b$ and $h$ are the real parameters. In the
distribution, the loss cone is empty for dimensionless angular
momentum $\alpha < h$. Tremaine studied the most large-scale
perturbations with the spherical indices $l=1$ and $l=2$.

In this section we consider two monoenergetic models. The first
one is
\begin{align}\label{eq:3.2}
f(\alpha)= N\,\ln\,\Bigl(\frac{\alpha^2}{h^2}\Bigr), \ \
h<\alpha<1,
\end{align}
where $h<1$ characterize size of the loss cone, $N$ is the
normalization constant. Dependence of distributions (\ref{eq:3.1})
and (\ref{eq:1.1}) with $f(L)$ from (\ref{eq:3.2}) on the angular
momentum is identical. Just this dependence is crucial for
stability or instability of each specific distribution. Stability
of distribution (\ref{eq:3.2}) for arbitrary values of $l$ is
proved in  Sec. 2.4.1.

Another distribution (Sec. 2.4.2) is the simplest {\it monotonic
Heaviside} model in a form of the step-like DF,
\begin{align}\label{eq:3.4}
f(\alpha)= \frac2{1-h^2}\,H(\alpha-h)\,H(1-\alpha).
\end{align}
The factor $H(1-\alpha)$ is added to reflect that the DF domain is bounded by
circular orbits.

In Sec. 2.4.3, we prove the stability of spherical systems (in the
field of a central massive body), all orbits of which are
circular.

\subsubsection{The log model}

Distribution in the form (\ref{eq:3.2}) allows to calculate the
density $\rho(r)$ explicitly. Using (\ref{eq:1s.10}) one obtains:
\begin{align}\label{eq:3.5}
    {\bar \rho}\,(r)= N \il_{h^2}^{\alpha_{\max}^2(r)}
    \frac{d\,(\alpha^2)}{\sqrt{\alpha_{\max}^2(r) -\alpha^2 \phantom{\big|}}} \, \ln
    \Bigl(\frac{\alpha^2}{h^2}\Bigr),
\end{align}
where the normalization constant satisfies the relation:
$$
N^{-1} = \il_h^1 d\alpha\, \alpha \ln \(\frac{\alpha^2}{h^2}\) =
\frac 12 \Bigl[\,\ln\,\Bigl(\frac{1}{h^2}\Bigr)-(1-h^2)\Bigr].
$$
From the condition $h^2 \le \alpha_{\rm max}^2(r)\equiv
({4\,r}/{R})\,\left(1-{r}/{R}\right)$, it follows that
$$
 R_{\min} = \ccase{1}{2}\,R\,(1-\sqrt{1-h^2}),\qquad R_{\max} =
 \ccase{1}{2}\,R\,(1+\sqrt{1-h^2}).
$$
In presence of the loss cone, $h>0$, the radius of the system
$R_{\rm max}$ is less than the apocenter radius for a radial orbit,
$R$, since stars with low angular momentum, $\alpha<h$, are
absorbed by the black hole. Integration (\ref{eq:3.5}) gives for
the density
\begin{multline*}
  \bar \rho\,(r) = \frac 4R \left[\sqrt{r\,(R-r)}\right.\times\\
  \times\ln\frac{\sqrt{r\,(R - r)} + \sqrt{(r - R_{\rm min})
  (R_{\rm max} - r)}}
  {\sqrt{R_{\rm min}\,R_{\rm max}\phantom{\big|}}}-\\
  -\left.\sqrt{(r-R_{\rm min}) (R_{\rm max}-r)}\right].
\end{multline*}
As is seen the density vanishes smoothly at the boundaries of the
spherical layer $r=R_{\rm min}$ and $r=R_{\rm max}$. The
expressions for the precession velocity are defined by the
formulas (\ref{eq:1s.8})\,--\,(\ref{eq:1s.9}).

\begin{figure}
 \centerline{\includegraphics[width=100mm]{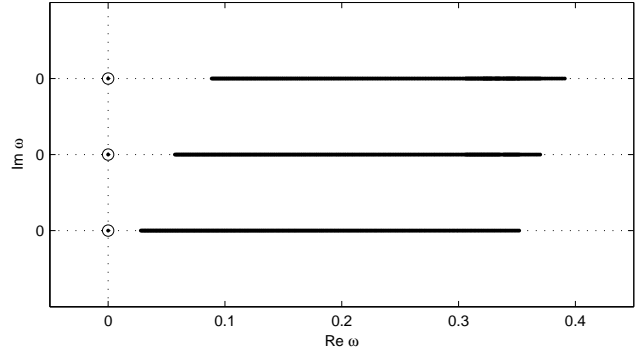}}
\caption{Spectrum for the log model with $h=0.1,\, 0.2,\, 0.3$ for $l=1$. In all
cases the eigenfrequencies are neutral and consist of one zero mode (circled
point) and continuous part of spectrum (points run into a line). To save room the
spectra are shown in a single plot, but vertically separated from one another
($h$ grows from bottom to top).}
 \label{fig:3}
\end{figure}

\bigskip
In Fig.\,\ref{fig:3}, the frequency spectrum, $\bar\omega$, of the spherical
harmonic $l=1$ for the {\it log}\,-model is presented for different values of the
parameter $h$. All calculations detect zero modes.

For other values of $l$ (we considered $l=2,\,3$), the discrete
modes are absent in the frequency spectra for all $h$.
The spectra are continuous and lie at the region of real and
positive values of ${\bar\omega}^2$. We conclude that the log
models are turned out to be stable.

\subsubsection{The monotonic Heaviside model}

Following the procedure described in Sec. 2.3.2 for the unstable
Heaviside model, one can derive the following equation for the distribution function (\ref{eq:3.4}):
\begin{align}\label{eqv:5.6}
    \phi_{s}(h)
    =-\frac{C_l}{C_1}\,\frac{2}{\sqrt{1-h^2\phantom{\big|}}}\!\!\!\sum\limits_{s'=s_{\rm
    min}}^l\!\!\!\!\!\,D_{l}^{s'}
    \dfrac{s'^{\,2}} {\lambda^2-s'^{\,2}}\,{\cal
    K}_{\,s\,s'}^{(l)}(h,h)\,\phi_{\,s'}(h),
\end{align}
where $\lambda={\bar\omega}/{\nu(h)}$, $\nu(h)=8h/[3\,\pi^2 e(h)]$. It is easy to
see that for $l=1$ there is a zero mode only, since ${\cal
K}_{11}^{(1)}(h,h)=e(h)=(1-h^2)^{1/2}$. Introducing new variables $
X_s={s\,\phi_s(h)\sqrt{D_l^s\phantom{\big|}}}/(\lambda^2-s^2)$ one can reduce
(\ref{eqv:5.6}) to a standard linear set
\begin{align}\label{eqv:5.7}
    \lambda^2 X_s = s^2 X_s -\sum\limits_{s'=s'_{\rm min}}^l {\hat
    L}_{s\,s'}^{(l)}(h)\,X_s',
\end{align}
where the matrix
$$
{\hat L}_{s\,s'}^{(l)}(h)=2\,\frac{C_l}{C_1}\,\frac{{\cal
K}_{s\,s'}^{(l)}(h,h)}{e(h)}\,s\,s'\,
 \sqrt{D_l^s\,D_l^{s'}\phantom{\big|}}
$$
is Hermitian. So, the eigenfrequencies $\lambda^2$ are real. Here
again we have a competition of opposite factors, expressed by the
first and the second terms in the r.h.s of (\ref{eqv:5.7}). To
conclude whether the instability occurs, we must find numerically
zeros of the determinant
$\bigl|\!\bigl|(s^2-\lambda^2)\,\delta_{s\,s'}-{\hat
 L}^{(l)}_{s\,s'}\bigr|\!{\bigr|}=0,
$ as a function of $\lambda^2$ for $l\ge 2$. A rank of the
determinant is equal to $\bigl[\frac{1}{2}\,(l+1)\bigr]$.

\begin{figure}
 \centerline{\includegraphics[width=90mm]{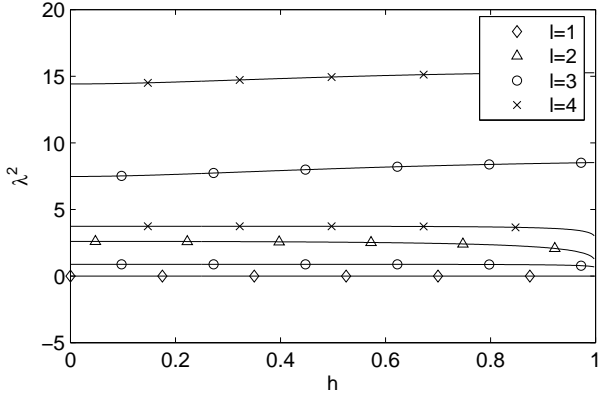}}
\caption{The dependence of the eigenfrequencies squared,
$\lambda_j^2(h)$, for $l=1, 2, 3, 4$.}
 \label{fig:4}
\end{figure}

\begin{figure}
 \centerline{\includegraphics[width=90mm]{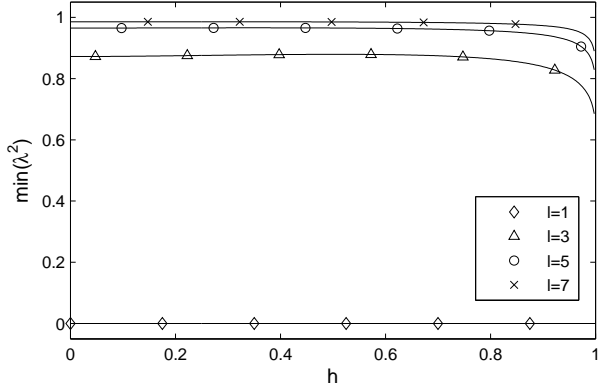}}
\caption{The dependence of the smallest (for given $l$)
eigenfrequencies squared, $\lambda^2(h)$, for $l=1,\ 3,\ 5,$ and
7.}
 \label{fig:5}
\end{figure}

The results are represented in Figs. \,\ref{fig:4} and
\ref{fig:5}. This figures show the dependence of $\lambda_j^2(h)$,
$\ \ j=1,2,..., [\frac{1}{2}\,(l+1)]$ as a function of the loss
cone size $h$. The instability is clearly absent. Each eigenvalue
$\lambda_i^2(h)$ has only a weak dependence on $h$ and
approximately equals to $j^2$. The least stable mode is $l=3$ (see
Fig. \,\ref{fig:5}), but it is still far from instability.

\subsubsection{Model with circular orbits}

Let us consider the simplest monotonic model, in which all orbits are circular.
In this section we do not assume a distribution to be monoenergetic, otherwise the
density distribution would be degenerated to a thin spherical layer.
Let us assume that the DF is
 $F(E,L)=A\, \delta\,\bigl[\,L-L_{\rm circ}(E)\bigr]$, where $A={\rm const}$
is normalization factor and  $E$  changes in some range $\Delta
E$. In terms of radial and transverse velocities $v_r$ and
$v_{\perp}=\sqrt{v_\theta^2+v_\varphi^2}$, the DF is:
  $$
  F(v_r,v_{\perp},r)=\frac{\rho_0(r)}{2\pi\,
  v_0(r)}\,\,\delta(v_r)\,\delta[v_{\perp}-v_0(r)], \ \ v_0(r)=r\Omega(r),
  $$
where $\Omega(r)$ is the angular velocity of a star on the
circular orbits. This velocity is determined by a balance of the
centrifugal force and a sum of the gravitational forces from the
central body and from the spherical cluster:
 $\Omega^2=\Omega_0^2(r)+({1}/{r})\,{d\Phi_G(r)}/{dr},\ \
\Omega_0^2(r)={GM_c}/{r^3}. $ Here we also suggest that
$\Omega_0^2\gg r^{-1}\,{d\Phi_G(r)}/{dr}. $

In this approximation, the orbits are near-Keplerian, and the
following relations are valid:
\begin{align}\label{eqv:5.8}
 \omega_0^2&\equiv 4\pi
 G\rho_0(r)=\Phi_G''+\frac{2}{r}\,\Phi_G',\nonumber\\
 \Omega&=\Omega_0+\frac{1}{2r\Omega_0}\,\Phi_G',\\
 \kappa&=\Omega_0+\frac{1}{2\Omega_0}\Bigl(\Phi_G''+\frac{3}{r}\,\Phi_G'\Bigr).\nonumber
\end{align}
 For the precession rate we have (see also Tremaine, 2001)
 $ \Omega_{\rm
 pr}=\Omega-\kappa=-({1}/{2\Omega_0})\,\left[\Phi_G''+({2}/{r})\,\Phi_G'\right],$
 or, taking into account (\ref{eqv:5.8}),
 $
 \Omega_{\rm pr}=-\frac{1}{2}\,{\omega_0^2}/{\Omega_0}.
 $
Since $\epsilon = M_G/M_c\ll 1$, one has the following scalings: $\Omega_{\rm pr}
\sim \epsilon\,\Omega_0$, $\omega_0^2 \sim \epsilon\,\Omega_0^2$, and for slow
modes $\omega\sim \Omega_{\rm pr}\sim \epsilon\,\Omega_0$.

We start from the equation derived by Pal'chik et. al. (1970), for
the models with circular orbits (the equation can also be found in
the monograph by Fridman and Polyachenko (1984).\footnote{Note
that both in the monograph and in the original paper, the form of
equation does not allow to include the external gravitational
field from a halo or a central body. We have slightly changed the
equation to make it possible.} It has a form:
\begin{align}\label{eqv:5.9}
   \frac{d}{dr}\,r^2 A_l(r,\omega)\frac{d\chi_l}{dr}-B_l(r,\omega)\,\chi_l(r)=0,
\end{align}
where $\chi_l(r)$ is a radial part of the potential perturbation
$\Phi(r)\propto
\chi_l(r)\,Y_l^m(\theta,\varphi)\,\exp\,(-i\,\omega t)$,
coefficients $A_l(r,\omega)$ and $B_l(r,\omega)$ are
\begin{align}\label{eqv:5.10}
  A_l(r,\omega)=1\!+\!\omega_0^2\!\sum\limits_{s=-l}^l\frac{D_l^s}
  {[\omega\!-\!(s\Omega\!-\!\kappa)][\omega\!-\!(s\Omega\!+\!\kappa)]},
\end{align}
\begin{multline}\label{eqv:5.11}
  B_l(r,\omega)=l\,(l+1)+\sum\limits_{s\,=-l}^l D_l^s\times\\
  \left\{r^2\frac{d}{dr}\left[
  \frac{\omega_0^2}{r}\,\frac{2s\Omega}
  {(\omega-s\Omega)\,[\omega-(s\Omega-\kappa)][\omega-
  (s\Omega+\kappa)]}\right]+
   \right.\\
    +\omega_0^2\,\left[\frac{4s\Omega(\omega-s\Omega)+
   s^2\bigl[(\omega-s\Omega)^2+4\Omega^2-\kappa^2\bigr]}
   {(\omega-s\Omega)^2\,
   [\omega-(s\Omega-\kappa)][\omega-
  (s\Omega+\kappa)]}+\right.\\
   \left.\left.
   \frac{(l+s+1)(l-s)}{(\omega-s\Omega)\,(\omega-s\Omega-2\Omega)}\right]\right\}.
\end{multline}

Now we need to distinguish between even and odd values of $l$,
since $l$ and $s$ should be of the same parity (i.e. both even or
both odd).

For {\it even} $l$, the dominating contributions is expected from
$s=0$ and $s=-2$. However, one can see that for even $l$, the
contributions from $s=0$ and $s=-2$ cancel each other. Indeed,
setting $\omega_0^2=-2\Omega_0\Omega_{\rm pr}$ one has:
$$
A_l=1,\ \
 B_l=l(l+1)+D_l^0\,l\,(l+1)\,\frac{\Omega_{\rm
pr}}{\Omega}-D_l^2\,(l-1)\,(l+2)\frac{\Omega_{\rm pr}}{\Omega}.
$$
After taking into account the relation \[D_l^2=D_l^0
\bigl\{l\,(l+1)/[(l-1)(l+2)]\bigr\},\] one obtains: $A_l=1, \ \
B_l=l(l+1)$. So the equation (\ref{eqv:5.9}) is reduced to the
trivial relation $\Delta\chi_l=0$, which means the absence of the
slow density perturbations.

For {\it odd} $l$ the terms $s=\pm 1$ give the main contribution
to the sum, while other terms ($|s|\ne 1$) are beyond the accuracy
of the slow mode equation. Thus, one has:
\begin{align}\label{eqv:5.12}
A_l&=1+2\,D_l^1\,\frac{\Omega_{\rm pr}^2}{\omega^2-\Omega_{\rm
pr}^2}, \nonumber\\
B_l&=l\,(l+1)-4\,D_l^1\,r^2\frac{d}{dr}\,\left(\frac{1}{r}\, \frac{\Omega_{\rm
pr}^2}{\omega^2-\Omega_{\rm pr}^2}\right).
\end{align}
To study this case we transform the differential equation
(\ref{eqv:5.9}) with $A_l$ and $B_l$ from (\ref{eqv:5.12}) to an
integral equation. Eq.\,(\ref{eqv:5.9}) can be represented in the
form of the Poisson equation
\begin{align}\label{eqv:5.13}
\Delta \chi_l(r)=4\pi G\rho_l(r),
\end{align}
with the perturbed density
\begin{align}\label{eqv:5.14}
\rho_l(r)&=-\frac{D_l^1}{4\pi G}\,\Biggl\{\frac{1}{r^2}\,\frac{d}{dr}\Bigl[\,r^2
S(\omega^2,r)\,\frac{d\chi_l}{dr}\Bigr]\Biggr.\nonumber\\
\Biggl. &+\frac{d}{dr}
\Bigl[\,\frac{2}{r}\,S(\omega^2,r)\Bigr]\,\chi_l\Biggr\},
\end{align}
where
 $S(\omega^2,r)={2\,\Omega_{\rm pr}^2}/({\omega^2-\Omega_{\rm pr}^2})$.
The solution of Eq. (\ref{eqv:5.13}) in the integral form is:
\begin{align}\label{eqv:5.15}
 \chi_l(r)=-\frac{4\pi G}{2l+1}\int r'^{\,2} dr'\,\rho_l(r')\,{\cal F}_l(r,r')
\end{align}
where the kernel
\begin{align}\label{eqv:5.16}
 {\cal
 F}_l(r,r')=\frac{(r')^l}{r^{l+1}}\,H(r-r')+\frac{r^l}{(r')^{l+1}}\,H(r'-r),
\end{align}
or, substituting (\ref{eqv:5.14}) into (\ref{eqv:5.15}), and
integrating by parts,
\begin{multline}\label{eqv:5.17}
\chi_l(r)=-\frac{D_l^1}{2l+1}\int dr'
S(\omega^2,r')\,\frac{d}{d\,r'}\,[r'^{\,2}\chi_l(r')]\,\times\\
\times\left[\frac{d\,{\cal F}_l(r,r')}{dr'}+\frac{2}{r'}\,{\cal
F}_l(r,r')\right].
\end{multline}
Applying the operator $ {\hat{\cal P}}(r)={d}/{dr}+{2}/{r} $ to
both parts of Eq. (\ref{eqv:5.17}) and denoting
 $
 \Psi_l(r)={\hat{\cal
 P}}(r)\,\chi_l(r)={d\chi_l(r)}/{d\,r}+({2}/{r})\,\chi_l(r)
 $
we obtain an integral equation\footnote{ The integral equation
(\ref{eqv:5.18}) can also be derived from the general ``slow''
integral equation, by considering of circular orbit limit.
However, that derivation is much more cumbersome than one given
here.}
\begin{align}\label{eqv:5.18}
   \Psi_l(r)=-\frac{D_l^1}{2l+1}\int r'^2 dr' S(\omega^2,r')\,{\cal
   R}_l(r,r')\,\Psi_l(r'),
\end{align}
with the new symmetrical kernel
    $ {\cal R}_l(r,r')=
    \left( {d}/{dr}+{2}/{r}\right)\,\left(
    {d}/{dr'}+{2}/{r'}\right)\,{\cal F}_l(r,r'). $
Introducing the new function
 $
 Z_l={r\,\Omega_{\rm pr}}/({\omega^2-\Omega_{\rm pr}^2})\,\Psi_l,
 $
we obtain the required integral equation
\begin{multline}\label{eqv:5.19}
 [\omega^2-\Omega_{\rm pr}(r)^2]\,Z_l(r)=\\
 -\frac{2D_l^1}{2l+1}
 \int dr'\,\Omega_{\rm pr}(r)\,\Omega_{\rm pr}(r')\,{\cal K }_l(r,r')\,Z_l(r'),
\end{multline}
with the kernel  ${\cal K}_l(r,r')=r\,r'\,{\cal R}_l(r,r')$.

Since the kernel defines a self-adjoint integral operator, all
eigenfrequencies $\omega^2$ should be real. To determine whether
negative values of $\omega^2$ are possible, let us write out the
kernel ${\cal K}_l(r,r')$ explicitly:
\begin{align}\label{eqv:5.20}
{\cal K}_l(r,r')=-(l+2)\,(l-1)\,{\cal
F}_l(r,r')+(2\,l+1)\,\delta(r-r').
\end{align}
It contains two contributions: the first is negative, the second
is positive. Substituting (\ref{eqv:5.20}) into (\ref{eqv:5.19}),
one finds
\begin{multline}\label{eqv:5.21}
 \omega^2 Z_l(r)=\Omega_{\rm pr}^2(r)(1-2D_l^1)\,Z_l(r)+
 2\,D_l^1\,\frac{(l+2)\,(l-1)}{2\,l+1}\times\\
 \times\int d\,r'\, \Omega_{\rm pr}(r)\,\Omega_{\rm pr}(r')\,{\cal F }_l(r,r')\,Z_l(r').
\end{multline}

For $l=1$ one can see that $\omega^2=0$ satisfies this equation
since $D_1^1=\case{1}{2}$. However, the most interesting fact
consists in stability of all higher modes, $l\ge 3$. Indeed, since
$1-2D_1^l>0$ for $l\ge 3$, and the integral operator in the r.h.s.
is positively defined, we conclude that all eigenvalues,
$\omega^2$, are positive. Consequently, the instability is absent
in the limit of circular orbits. The result is universal and does
not depend on a particular choice of the model $\Omega_{\rm pr}$.

\section{Thin disk systems}

\subsection{Slow mode Integral equation for monoenergetic disk models}

In this section we shall consider the monoenergetic distributions
of  (\ref{eq:1.1}) type assuming the function $f(\alpha)$ to be
even, $f(\alpha) = f(-\alpha)$. The function $F(E,L)$ is
normalized as follows:
\begin{align}\label{eq:disk.2}
    M_G = \int F\, d\,\Gamma=(2\pi)^2\int\frac{dE}{\Omega_1(E)}
    \int\limits_{-L_{\rm circ}(E)}^{L_{\rm circ}(E)} d\,L\,F(E,L),\ \
\end{align}
which gives the normalization constant
\begin{align}\label{eq:disk.3}
    A = \frac{M_G}{\pi^2 R^2}
\end{align}
provided that $ \int_{-1}^1 f(\alpha)\,d\alpha=1$.

The integral equation for slow modes (Paper I) can be represented
in the form:
\begin{align}\label{eq:disk.4}
   \phi(\alpha)=\frac{C_m}{\pi^3}\int\limits_{-1}^1
    \frac{d\alpha'\,d{f}/d\alpha'}
    {\bar \omega - \nu(\alpha')}\,\,
    {\cal K}_m(\alpha,\alpha')\,{\phi}(\alpha'),
\end{align}
or, using the evenness of $\phi(\alpha)$ which stems from evenness
of $f(\alpha)$, oddness of $\Omega_{\rm pr}(\alpha)$, and symmetry
properties of the kernel, ${\cal K}_m(\alpha,\alpha') = {\cal
K}_m(-\alpha,\alpha')$,
\begin{align}\label{eq:disk.4a}
   \phi(\alpha)=\frac{2\,C_m}{\pi^3}\int\limits_{0}^1
    \frac{\nu(\alpha')}{{\bar \omega}^2 - \nu^2(\alpha')}\,\frac{df}{d\alpha'}
    \,\,
    {\cal K}_m(\alpha,\alpha')\,{\phi}(\alpha')\,d\alpha',
\end{align}
where $\bar \omega$ and $\nu(\alpha)$ are the dimensionless
pattern speed and the dimensionless precession rate:
\begin{align}\label{eq:disk.4s2}
    \bar\omega=\frac{\Omega_p}{\epsilon\,\Omega_1},\ \ \
    \nu (\alpha)=\frac{\Omega_{\rm
    pr}(\alpha)}{\epsilon\,\Omega_1}.
\end{align}

Changing the unknown function, integral equation
(\ref{eq:disk.4a}) takes the form of linear eigenvalue problem:
\begin{align}\label{eq:disk.4s}
 \bigl[\,{\bar\omega}^2-\nu^2(\alpha)\bigr]\,\psi(\alpha)=
 \frac{2\,C_m}{\pi^3}
  \int\limits_{0}^1
   \nu(\alpha')\,\frac{df}{d\alpha'}\,
    {\cal K}_m(\alpha,\alpha')\,{\psi}(\alpha')\,d\alpha'.
    \end{align}
The kernel functions for thin disks ${\cal K}_m(\alpha,\alpha')$
can be transformed from the corresponding expression in Paper I to
a suitable form as follows:
\begin{align}\label{eq:disk.5}
    {\cal
    K}_m(\alpha,\alpha')=\frac{1}{C_m}\int\limits_{0}^{\pi}d\tau\,
    r\,\cos m\zeta \int\limits_{0}^{\pi}d\tau'\, r'\,\cos
    m\zeta'\,\, {\cal F}_m(r, r'),
\end{align}
where dependence of $r$ and anomaly $\zeta$ on $\tau$ and $e$ are the same as in
the spherical case (\ref{eq:1s.6}), but the function ${\cal F}_m(x,y)$ is
\begin{align}\label{eq:disk.6}
{\cal F}_m(x,y)=\int\limits_{-\pi}^{\pi}\frac{\cos m\theta\,d\theta}{\sqrt{
x^2+y^2-2xy\cos\theta\phantom{\big|}}}.
\end{align}
As before, the kernel ${\cal K}_m(\alpha,\alpha')$ is normalized
to unity: ${\cal K}_m(0,0)=1$, which means $C_m$ equal to
\begin{align}\label{eq:disk.7}
    C_m=\int\limits_0^1 dx\,\sqrt{\frac{x}{1-x}\phantom{\bigg|}}
    \int\limits_0^1 dy\,\sqrt{\frac{y}{1-y}\phantom{\bigg|}}\,\,{\cal
    F}_m(x,y).
\end{align}
The latter formula immediately follows from (\ref{eq:disk.5}) if
one reminds that for radial orbits $\zeta=\pi$,\ \ \ $\cos
m\zeta=(-1)^m,$\ \ \ $d\tau= dx\,[\,x\,(1-x)]^{-1/2}$. For the
lowest azimuthal numbers, functions ${\cal F}_m(x,y)$ can be
expressed through elliptical integrals of the first and the second
kind ${\bf K}(q)$ and ${\bf E}(q)$:
\begin{align}\label{eq:disk.8}
    {\cal F}_1(x,y)= \frac{4}{r_>}\,\frac{{\bf K}(q)-{\bf
    E}(q)}{q},
    \end{align}
    \begin{align}\label{eq:disk.9}
    {\cal F}_2(x,y) &= \frac{4}{3r_>}\,\left[\left(\frac{2}{q^2}+1\right){\bf
    K}(q)-2\,\left(\frac{1}{q^2}+1\right){\bf E}(q)\right],
\end{align}
where $r_>={\rm max}(x,y)$,  $r_<={\rm min}(x,y)$,
$q={r_<}/{r_>}$. Using (\ref{eq:disk.8}) and (\ref{eq:disk.9}) one
can obtain numerically $C_1=10.88$, $C_2=7.45$.
\medskip

For the surface density we have:
\begin{align}\label{eq:disk.10}
\sigma_0(r)=\frac{2}{r}\int dE\int\limits_{-L_{\rm max}(r)}^{L_{\rm
max}(r)}\frac{F(E,L)\,dL}{\sqrt{2E+\dfrac{2GM_c}{r}-\dfrac{L^2}{r^2}\phantom{\Big|}}}=
\frac{2M_G}{\pi^2 R^2}\, \Sigma_0(r),
\end{align}
where
\begin{align}\label{eq:disk.11}
\Sigma_0(r)=\int\limits_{-\alpha_{\rm max}(r)}^{\alpha_{\rm
max}(r)}\frac{f(\alpha)\,d\alpha}{\sqrt{\alpha_{\rm
max}^2(r)-\alpha^2\phantom{\big|}}},
\end{align}
and $\alpha^2_{\rm max}(r)=L^2_{\rm max}(r)/L^2_{\rm
circ}=4\,(r/R)\,(1-r/R)$ as for spheres.
\medskip

The relation between the precession rate and the potential
$\Phi_G(r)$ is the same as in the spherical systems, (see Tremaine
2005 and Paper I)
\begin{align}\label{eq:disk.12}
    \Omega_{\rm pr}=\frac{8}{\pi}\frac{1}{e\,\Omega_1
    R^3}\int\limits_0^{\pi}
    r^2\,\frac{d\Phi_G}{d\,r}\,\cos\zeta\,d\zeta.
\end{align}
but the relation between the potential and the surface density is
much more complicated\footnote{For that reason the precession in
the near-Keplerian disk is not always retrograde.}
 (see, e.g., Tremaine, 2001)
\begin{align}\label{eq:disk.3.16}
\Phi_G(r)=-\frac{4G}{r^{1/2}}\int
(r')^{1/2}\,\sigma_0(r')\,d\,r'\,\bigl[\,q^{1/2}\,{\bf K}(q)
 \bigr].
 \end{align}
Using (\ref{eq:disk.10})\,--\,(\ref{eq:disk.3.16}) one obtains a
suitable expression for the scaled precession rate $\nu(\alpha)$
(\ref{eq:disk.4s2}) in the integral form
\begin{align}\label{eq:disk 3.17}
\nu(\alpha)={\alpha}\int\limits_0^1 d\alpha'\,
 f(\alpha')\,{\cal Q}(\alpha,\alpha'),
   \end{align}
where ${\cal Q}(\alpha,\alpha')$ is a universal function (i.e.
does not depend on form of the distribution):
$$   {\cal Q}(\alpha,\alpha')=\frac{1}{\pi^3 e^2}
   \int\limits_{r'_{\rm min}}^{r'_{\rm max}}
 \frac{r'\,dr'}{\sqrt{(r'-r'_{\rm min})(r'_{\rm max}-r')\phantom{\big|}}}\times
$$
 \begin{align}\label{eq:disk 3.18}
 \times\!\!\!\!
  \int\limits_{r_{\rm min}}^{r_{\rm max}}\!\! dr\,
 \frac{2r-\alpha^2}{r\,\sqrt{(r\!-r_{\rm min})(r_{\rm max}\!-r)\phantom{\big|}}}\,
 \Biggl[\,\frac{{\bf
E}\,(\kappa)}{r'-r}-\frac{{\bf K}\,(\kappa)}{r'+r}\Biggr].
   \end{align}
Here $\kappa=2\,\sqrt{\,r\,r'}/(r+r')$. The integral of the first
term is understood in the principle value sense. Using the same
trick as in Sec. 2, one can change to new integrating variables
$\tau$ and $\tau'$, where $r=\frac{1}{2}\,(1-e\,\cos\tau)$ and
$r'=\frac{1}{2}\,(1-e'\,\cos\tau')$. Then, for ${\cal
Q}(\alpha,\alpha')$ one obtains
\begin{multline}\label{eq:disk 3.19}
{\cal Q}(\alpha,\alpha')=\dfrac{1}{2\,\pi^3\,e}
  \int\limits_{0}^{\pi} d\tau\,
 \dfrac{e-\cos\tau}{1-e\cos\tau}
  \int\limits_{0}^{\pi} d\tau'\,(1-e'\cos\tau')\,\times\\
  \times
 \Biggl[\,\dfrac{{\bf
E}\,(\kappa)}{r'-r}- \dfrac{{\bf K}\,(\kappa)}{r'+r}\Biggr].
  \end{multline}

\subsection{Variational principle and sufficient condition for instability of
$m=1$ mode}

As we see from Sec. 2, for spherical systems with the monotonic DF,
the variational principle takes place. Besides, for $l=1$ and the
empty loss cone, a zero frequency solution exists which stands for
a sphere displacement from the massive center; all other
eigenmodes being stable. A thin disk is completely different. The
displacement is no longer an eigenmode. Moreover, models with
analogous distributions are turn out to be unstable. Let us prove
the instability of lopsided $m=1$ mode provided that
\begin{align}\label{eq:disk 3.20}
g(\alpha)=\nu(\alpha)\,{df}/{d\alpha}<0.
\end{align}
Note that spherical models with the analogous condition are
stable.

Disks with even DFs satisfying condition (\ref{eq:disk 3.20}) obey
the variational principle, which means the eigenfrequencies
squared are real. So one can formulate a {\it sufficient}
condition of instability for $m=1$ azimuthal perturbations as
follows: If the loss cone is empty ($f(0)=0$), the DF is
monotonically increasing $df/d|\alpha|>0$, and the precession is
retrograde for all values of angular momentum
($\nu(\alpha)/\alpha<0$), then $m=1$ perturbations are unstable.

To prove the statement we shall use integral equation
(\ref{eq:disk.4a}), in which $\bar\omega = i\Gamma$ with
$\Gamma>0$ is assumed:
\begin{align}\label{eq:disk 3.21}
    {\cal M}(\Gamma)\,\phi(a)=0,
\end{align}
where operator ${\cal M}(\Gamma)$ is
\begin{align}\label{eq:disk 3.22}
 {\cal M}(\Gamma)\,\phi(a)\equiv\phi(a)+\frac{2\,C_m}{\pi^3}\int\limits_{0}^1
    \frac{g(\alpha')}
    {\Gamma^2+\nu^2(\alpha')}\,
    {\cal K}_m(\alpha,\alpha')\,{\phi}\,(\alpha')\, d\alpha'.
\end{align}
Now we consider another eigenvalue problem
\begin{align} \label{eq:disk 3.23}
    {\cal M}(\Gamma)\,\phi\,(a)=\lambda\,(\Gamma)\,\phi\,(a).
\end{align}
Eigenvalues $\Gamma$ of the problem (\ref{eq:disk 3.21})
correspond to eigenvalues $ \lambda\,(\Gamma)=0$ of the problem
(\ref{eq:disk 3.23}). Let us define an inner product as $ \langle
X, Y\rangle = \int_0^1 d\alpha\,X^{\ast}(\alpha)\,
W(\Gamma,\alpha)\, Y(\alpha),$ where the weight function
$W(\Gamma,\alpha)= -{g(\alpha)}/[\Gamma^2+\nu^2(\alpha)]>0.$
Operator ${\cal M}$ has the following properties.
\smallskip

1. ${\cal M}$ is Hermitian, i.e. $ \langle \psi(a), {\cal
M}\phi(a)\rangle=\langle {\cal M}\psi(a),\phi(a)\rangle= \langle
\phi(a),{\cal M}\psi(a)\rangle^{\ast}. $
\smallskip

2. ${\cal M}$ is continuous, when $\Gamma\ge 0$. One might think
that the first term $\phi(a)$ in the r.h.s. of (\ref{eq:disk
3.22}) breaks down the continuity, which in turn means that system
of proper functions is incomplete. However, this is not the case,
since $\phi(a)$  can be absorbed by introducing new eigenvalue
$\Lambda(\Gamma)=\lambda(\Gamma)-1$ in (\ref{eq:disk 3.23}). Since
$f(\alpha)$ is even, for smooth DF we have $f(0)=f'(0)=0$,
$f''(0)>0$. This condition guarantee the weight function
$W(\alpha)$ to be finite even for $\Gamma=0$, despite $\nu={\cal
O}(\alpha)$ at $\alpha\to 0$.
\smallskip

3. ${\cal M}$ is positive definite at sufficiently large $\Gamma$.
This is evident, since $W(\alpha) > 0$, and the second term in
(\ref{eq:disk 3.22}) becomes small at large $\Gamma$.
\medskip

>From the first two properties it follows that for fixed $\Gamma\ge
0$ eigenvalues $\lambda_n(\Gamma)$,\ \ $n=1,2,3,\, \ldots$ of
${\cal M}(\Gamma)$ are real, and the system of proper functions is
complete. The third property means that at large $\Gamma$ all
eigenvalues $\lambda_n(\Gamma)$ are positive.

\smallskip
If we find a test function $\phi_t(\Gamma_0,\alpha)$, for which the scalar
product $\langle\phi_t(\Gamma_0,\alpha),{\cal
M}(\Gamma_0)\,\phi_t(\Gamma_0,\alpha)\rangle $ is negative, it mean ${\cal
M}(\Gamma_0)$ is not positive definite for the given $\Gamma_0$. So, at least one
eigenvalue must be negative: $\lambda_{\rm min}(\Gamma_0)<0$. This minimal
eigenvalue $\lambda_{\rm min}(\Gamma)$ increases with $\Gamma$ and becomes
positive, as all other $\lambda_n(\Gamma)$.  We conclude that there must be a
value of $\Gamma$, $\Gamma_0<\Gamma<\infty$ for which $\lambda_{\rm
min}(\Gamma)=0$. This value is an eigenvalue for (\ref{eq:disk 3.22}), which
means the existence of the eigenmode describing aperiodic instability with growth
rate $\Gamma$.

For the test function $\phi_t(\Gamma_0,\alpha)$ one can take a
displacement of the disk from the center, which is similar to the
sphere displacement (\ref{eqv:2.15}) and correspond to the
lopsided perturbation $m=1$:
 $ \phi_t(\Gamma_0,\alpha)=({e}/\alpha)\,\,\nu(\alpha) $,
and $\Gamma_0=0$. One can show that
\begin{align} \label{eq:disk 3.24}
 \langle\phi_t(\Gamma_0,\alpha),{\cal M}(\Gamma_0)\,\phi_t(\Gamma_0,\alpha)\rangle<0.
\end{align}
Let the l.h.s. be $-P$, or, explicitly
\begin{multline*}
 P=\int\limits_0^1 d\alpha
 \frac{df(\alpha)}{d\alpha}\,\left[\frac{e(\alpha)}{\alpha}\right]^2\,
 \nu(\alpha)\,
 +\frac{2 C_1}{\pi^3}\int\limits_0^1 d\alpha
\,\frac{e(\alpha)}{\alpha}\times\\
\times\frac{df(\alpha)}{d\alpha}\,\int\limits_0^1 d\alpha' \,
\frac{e(\alpha')}{\alpha'}\,\frac{df(\alpha')}{d\alpha'}\,{\cal
K}_1(\alpha,\alpha').
 \end{multline*}
After some lengthy manipulations using (\ref{eq:disk.5}),
(\ref{eq:disk.8}), (\ref{eq:disk 3.19}), and condition $f(0)=0$,
one can show that $P$ is positive, so inequality (\ref{eq:disk
3.24}) is fulfilled.
\medskip

Tremaine (2005) has also obtained a sufficient condition for a
lopsided mode in the symmetrical disk using Goodman's (1988)
criterion. His condition, however, differs from ours. Namely: If
the loss cone is empty, $F(E,L=0)=0$, and $d\Phi_G(r)/dr>0$
throughout the radial range containing most of the disk mass, then
disk is unstable with respect to $m=1$ perturbations. This
formulation does not use the requirement for precession to be
retrograde and the DF to be monotonically increasing, although
monotonic increase of $F(E,0)=[\p F(E,L)/\p L]_{L=0}=0$, $[\p^2
F(E,L)/\p L^2]_{L=0}>0$ is implied, at least for small angular
momentum. Thus, the comparison between spherical and disk case can
hardly be made, unless the conditions of stability is formulated
in similar terms. To perform the comparison, we give our own
criterion that follows directly from the integral equation.

It needs to be emphasized that the sufficient condition by
Tremaine (2005) is different. Lack of the condition for the sign
of precession possibly means that his criterion include two type
of instability simultaneously~-- the radial orbit instability
arising in disks with prograde precession, and the loss cone
instability which require retrograde precession.

For disks composed of near-radial orbits, Tremaine's condition
gives the result obtained in Paper I: a disk with symmetrical
distribution $f(\alpha)$ obeying the conditions $f(0)=f'(0)=0$,
$f''(0)>0$ is unstable if the precession is retrograde. In turn,
precession of near-radial orbits is retrograde if
$d\Phi_G(r)/dr>0$.

\subsection{Numerical results}

To support the mathematical rationale given above and to provide a
basis for possible simulations, it is useful to obtain
eigenfrequencies of unstable modes for particular models. Here we
consider the {\it power\,--\,exp} model with symmetrical
distribution
\begin{align}\label{eq:disk_4.1}
f(\alpha) = N \frac{\alpha^2}{\alpha_T^3}
\exp(-\alpha^2/\alpha_T^2),
\end{align}
where the normalization constant is
$$
N^{-1} = 2\il_0^{1/\alpha_T^2} x^2\exp(-x^2)\, dx.
$$
For $\alpha_T\ll 1$ the constant $N=2/\sqrt{\pi}$. Distributions
become monotonic in the interval $[0,1]$ when $\alpha_T\ge 1$.
Note, that when $\alpha_T\gg 1$ DF is simply
$f(\alpha)=\frac{3}{2}\,\alpha^2$ in the interval $[-1,1]$ and
doesn't depend on $\alpha_T$.

Evaluation of integral equation (\ref{eq:disk.4s}) requires
preliminary calculations of the kernel function ${\cal
K}_m(\alpha,\alpha')$ using (\ref{eq:disk.5}), and the scaled
precession rate $\nu(\alpha)$ using (\ref{eq:disk 3.17}) and
(\ref{eq:disk 3.19}). For brevity, we skip the details here, just
noting that calculation of function $Q(\alpha,\alpha')$ is turn
out to be rather difficult numerical task. The calculations show
that for the model (\ref{eq:disk_4.1}) the precession rate
$\nu(\alpha)$ is retrograde for all $\alpha$, i.e.
$\nu(\alpha)/\alpha<0$.

\begin{figure}
 \centerline{\includegraphics[width=95mm, draft=false]{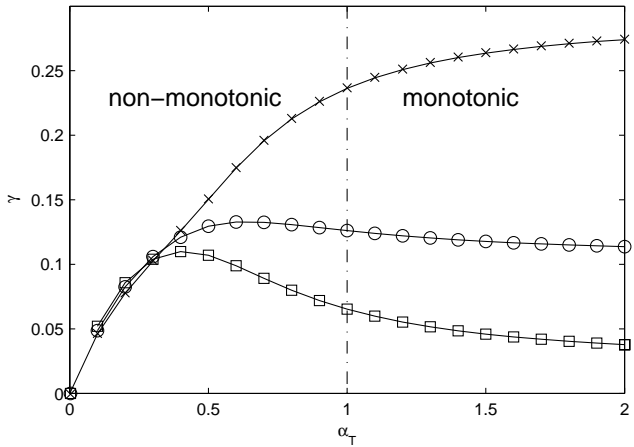}}
\caption{The dependence of $\gamma$ (growth rate divided by
azimuthal number $m$) vs. dimensionless angular momentum
dispersion $\alpha_T$ of the initial DF for azimuthal numbers $m=1$
(crosses), $m=2$ (circles), and $m=3$ (squares).} \label{fig:6}
\end{figure}

 The results
for values of azimuthal number $m=1,2,3$ are collected in Fig.\,\ref{fig:6}.
Since the initial distribution is symmetric, real parts of the eigenvalues
$\bar\omega$ are equal to zero. Hence in Fig.\,\ref{fig:6} we show the imaginary
parts $\gamma = \rmn{Im}\,\bar\omega$, which are the growth rates of the unstable
modes divided by azimuthal number $m$, vs. dimensionless angular momentum
dispersion $\alpha_T$. One can see that instability exists for all $\alpha_T$ and
never becomes saturated. Moreover, it is easy to obtain the asymptotic values
$\gamma$ for different $m$ at $\alpha_T\to\infty$: 0.289, 0.108, and 0.026 for
$m=1,2,3$ correspondingly. For small angular momentum, growth rates increase
linearly with $\alpha_T$, such as $\gamma/\alpha_T$ are equal to 0.454, 0.463,
and 0.481 for $m=1,2,3$ correspondingly.

\section{Discussion}

We have studied the stability of the spherically-symmetric and
thin disk stellar clusters around a massive black hole. We
conclude that stability properties of spherical clusters depend
crucially on monotonity of initial distribution functions, while
thin disk clusters are almost always unstable.

If the initial distribution of the spherical cluster is monotonic,
the cluster is most likely to be stable. This conclusion was first
made in Tremaine (2005), where stability of $l=1$ mode was
generally proved, and $l=2$ was tested numerically. We confirm
this conclusion by considering a number of monotonic distributions
for modes with arbitrary $l$. Besides, we have checked
distributions obtained from monotonic ones by making them vanish
quickly but smoothly at circular orbits. These models were also
stable. However, a general proof of stability for any monotonic
distributions was not yet found.

Spherical clusters with the non-monotonic DFs should be generally
affected by the gravitational loss-cone instability. The
instability was first found in our Paper I using a simplification
of systems with near-radial orbits. In the Sec. 2 we show that
this instability is due to just non-monotony of distributions over
angular momentum, the orbits may not necessary be near-radial.

In our opinion, both monotonic and non-monotonic distributions are
important for possible applications to real stellar clusters
around black holes. The DFs monotonically increasing from the loss
cone radius up to circular orbits are formed naturally due to
two-body collisions of stars. It follows from numerical
experiments (see, e.g., Cohn and Kulsrud, 1978), which predict
establishment of such distributions after a characteristic time
for collisional relaxation. These distributions may be
approximated by the formula $F\propto \ln \bigl(L/L_{\rm
min}\bigr)$.

Such a slowly increasing function is, in fact, predetermined by
the boundary conditions imposed in the cited numerical study and
some other investigations. Indeed, the vanishing condition at
$L=L_{\rm min}$, and the matching condition to isotropic
(Maxwellian) distribution, $F=F(E)$, at the boundary $E=E_{\rm
bound}=0$ of the phase space $(E,L)$ (boundary separates stars
which is gravitationally coupled to the black hole from the
others) is required. The last condition means the asymptotic (when
$E \to E_{\rm bound}$) independence of the function $F(E,L)$ on
the momentum $L$. So monotonic, or logarithmic, dependence of type
of (\ref{eq:3.2}) is quite reasonable.

The non-monotonic distributions are also real. If the cluster, is
formed, for example, as a result of the collisionless collapse
(several free fall times), then it remains collisionless for a
long timescale of collisional relaxation (see, e.g., Merritt \&
Wang, 2005). In principle, the system can have almost arbitrary DF
both in the energy and in the angular momentum. During the
collapse, a typical non-monotonic distribution of stars over the
angular momentum, with empty loss cone and maximum at some value
$L=L_{\ast}$, is formed.

In Paper I we argued that stability properties of such a
distribution is effectively analogous to one of typical plasma
distributions of the ``beam-like'' type. But they can readily
become unstable, as it is well-known in plasma physics (and also
confirmed by direct stability study of corresponding stellar
systems in Paper I). It is possible (as it is often so in plasma)
that for the time of collisionless behavior, DF can undergo a
dramatic change from its initial form. In particular, the
collective flux of stars into the loss cone caused by the
instability could, in principle, lead to the formation of a
considerable part of the black hole. Checking of such
possibilities is the most urgent task for future studies of
unstable {\it non-monotonic} models.

Since spherically-symmetric models with the {\it monotonic} DF are
apparently stable, but analogous disk systems are unstable (see
Tremaine 2005 and Sec. 3), a critical flatness of ellipsoid models
at which the instability begins is expected. Study of such
systems, as well as systems with more complex triaxial ellipsoids
can be performed using numerical simulations.

\section*{Acknowledgments}

We are grateful to V. A. Mazur for stimulating interest in our
work. The careful review of referee helped in improving the
presentation. The work was supported in part by Russian Science
Support Foundation, RFBR grants No. 05-02-17874, 08-02-00928 and
07-02-00931, ``Leading Scientific Schools'' Grants No.
7629.2006.2, 900.2008.2 and ``Young doctorate'' Grant No.
2010.2007.2 provided by the Ministry of Industry, Science, and
Technology of Russian Federation, and the ``Extensive objects in
the Universe'' Grant provided by the Russian Academy of Sciences,
and also by Programs of presidium of Russian Academy of Sciences
No 16 and OFN RAS No 16.

 \label{lastpage}

\end{document}